\documentclass[reprint,amssymb,superscriptaddress,nofootinbib]{revtex4-1}
\usepackage[pdftex]{graphicx}
\usepackage{subfig} 
\usepackage{color}
\usepackage{tikz}
\usetikzlibrary{shapes}
\definecolor{mycolor1}{rgb}{.9297,.5703,.6309}
\definecolor{mycolor2}{rgb}{.6212,.6114,.98575}
\definecolor{mycolor3}{rgb}{0.276,.776,.00}
\newcommand{\markerone}{\raisebox{1.8pt}{\tikz{\draw[color=mycolor1,ultra thick,dashed](0,1)--(0.55,1)}}}
\newcommand{\markertwo}{\raisebox{2pt}{\tikz{\draw[color=mycolor2,ultra thick,dashed](0,0.5)--(0.55,0.5)}}}
\newcommand{\markerthree}{\raisebox{1.9pt}{\tikz{\draw[color=mycolor3,ultra thick,dashed](0,0.5)--(0.55,0.5)}}}
\usepackage{tikz}
\usepackage{tikz-3dplot}
\usepackage{pgfplots}
\pgfplotsset{compat=newest}
\usepgfplotslibrary{smithchart}

\usetikzlibrary{
    arrows,
    positioning,
    calc,
    circuits.ee.IEC,
    external,
    plotmarks}

\usetikzlibrary{pgfplots.units}   
\usetikzlibrary{plotmarks}
\def\showpgfdiamond{\tikz[baseline=-0.9ex]\node[black,mark size=0.5ex,aspect=0.3,rotate=45]{\pgfuseplotmark{square*}};}
\def\showpgfdiamondnofill{\tikz[baseline=-0.9ex]\node[black,mark size=0.5ex,aspect=0.3,rotate=45,line width=1]{\pgfuseplotmark{square}};}
\usepackage[british]{babel}
\usepackage[autostyle]{csquotes}
\usepackage{amsmath,amsfonts,amssymb,mathtools}
\usepackage{diagbox}
\usepackage{bbold}
\usepackage{float}
\usepackage{float}
\usepackage{bbm,bm}
\usepackage[normalem]{ulem}
\usepackage{comment}
\usepackage{physics}
\usepackage{xcolor}

 \usepackage{multibib}
 \newcites{latex}{\LaTeX-Literature}

\usepackage[utf8]{inputenc}
\usepackage{pgfplots}
\definecolor{blueprl}{RGB}{46,48,146}
\usepgfplotslibrary{groupplots}
\newcommand{\I}{\mathrm{i}}
\usepackage{lipsum}

\def\ANU{Centre for Quantum Computation and Communication Technology, Department of Quantum Science, Australian National University, Canberra, ACT 2601, Australia.}
\def\NTU{School of Physical and Mathematical Sciences, Nanyang Technological University, Singapore 639673, Republic of Singapore}

  \def\Jena{Institute of Applied Physics, Abbe Center of Photonics, Friedrich-Schiller-Universität Jena, 07745 Jena, Germany}
  \def\Camb{Cavendish Laboratory, University of Cambridge, JJ Thomson Avenue, Cambridge CB3 0HE, United Kingdom}
  \def\Fraun{Fraunhofer-Institute for Applied Optics and Precision Engineering IOF, 07745 Jena, Germany}
  \def\MP{Max Planck School of Photonics, 07745 Jena, Germany}
  \def\Inns{Institut f$\ddot{\text{u}}$r Experimentalphysik, 6020 Innsbruck, Austria}
  \def\Sydney{School of Mathematical and Physical Sciences, Macquarie University, Sydney, NSW 2109, Australia}
   \def\Innsthree{Alpine Quantum Technologies (AQT), 6020 Innsbruck, Austria}
  \def\Innstwo{Institute for Quantum Optics and Quantum Information, 6020 Innsbruck, Austria}
  \def\AWS{Amazon Web Services, Canberra, ACT 2601, Australia}
  \def\Astar{Institute of Materials Research and Engineering, Agency for Science Technology and Research (A*STAR), 2 Fusionopolis Way, 08-03 Innovis 138634, Singapore}

\begin{document}
\title{Approaching optimal entangling collective measurements on quantum computing platforms}
\author{Lorc{\'a}n O. Conlon}
\email{lorcanconlon@gmail.com}
\affiliation{\ANU}
\author{Tobias Vogl}
\affiliation{\Jena}
\affiliation{\Camb}
\author{Christian D. Marciniak}
\affiliation{\Inns}
\author{Ivan Pogorelov}
\affiliation{\Inns}
\author{Simon K. Yung}
\affiliation{\ANU}
\author{Falk Eilenberger}
\affiliation{\Jena}
\affiliation{\Fraun}
\affiliation{\MP}
\author{Dominic W. Berry}
\affiliation{\Sydney}
\author{Fabiana S. Santana }
\affiliation{\AWS}
\author{Rainer Blatt}
\affiliation{\Inns}
\affiliation{\Innstwo}
\author{Thomas Monz}
\affiliation{\Inns}
\affiliation{\Innsthree}
\author{Ping Koy Lam}
\email{ping.lam@anu.edu.au}
\affiliation{\ANU}
\affiliation{\NTU}
\affiliation{\Astar}
\author{Syed M. Assad}
\email{cqtsma@gmail.com}
\affiliation{\ANU}
\affiliation{\NTU}

\begin{abstract}

Entanglement is a fundamental feature of quantum mechanics and holds great promise for enhancing metrology and communications. Much of the focus of quantum metrology to date has been on generating highly entangled quantum states which offer better sensitivity, per resource, than what can be achieved classically. However, to reach the ultimate limits in multi-parameter quantum metrology and quantum information processing tasks, collective measurements, which generate entanglement between multiple copies of the quantum state, are necessary. Here we experimentally demonstrate theoretically optimal single-copy and two-copy collective measurements for simultaneously estimating two non-commuting qubit rotations. This allows us to implement quantum-enhanced sensing, for which the metrological gain persists for high levels of decoherence, and to draw fundamental insights about the interpretation of the uncertainty principle. We implement our optimal measurements on superconducting, trapped-ion and photonic systems, providing an indication of how future quantum-enhanced sensing networks may look.

\end{abstract}
\maketitle

\begin{figure*}[t]
\includegraphics[width=0.95\textwidth]{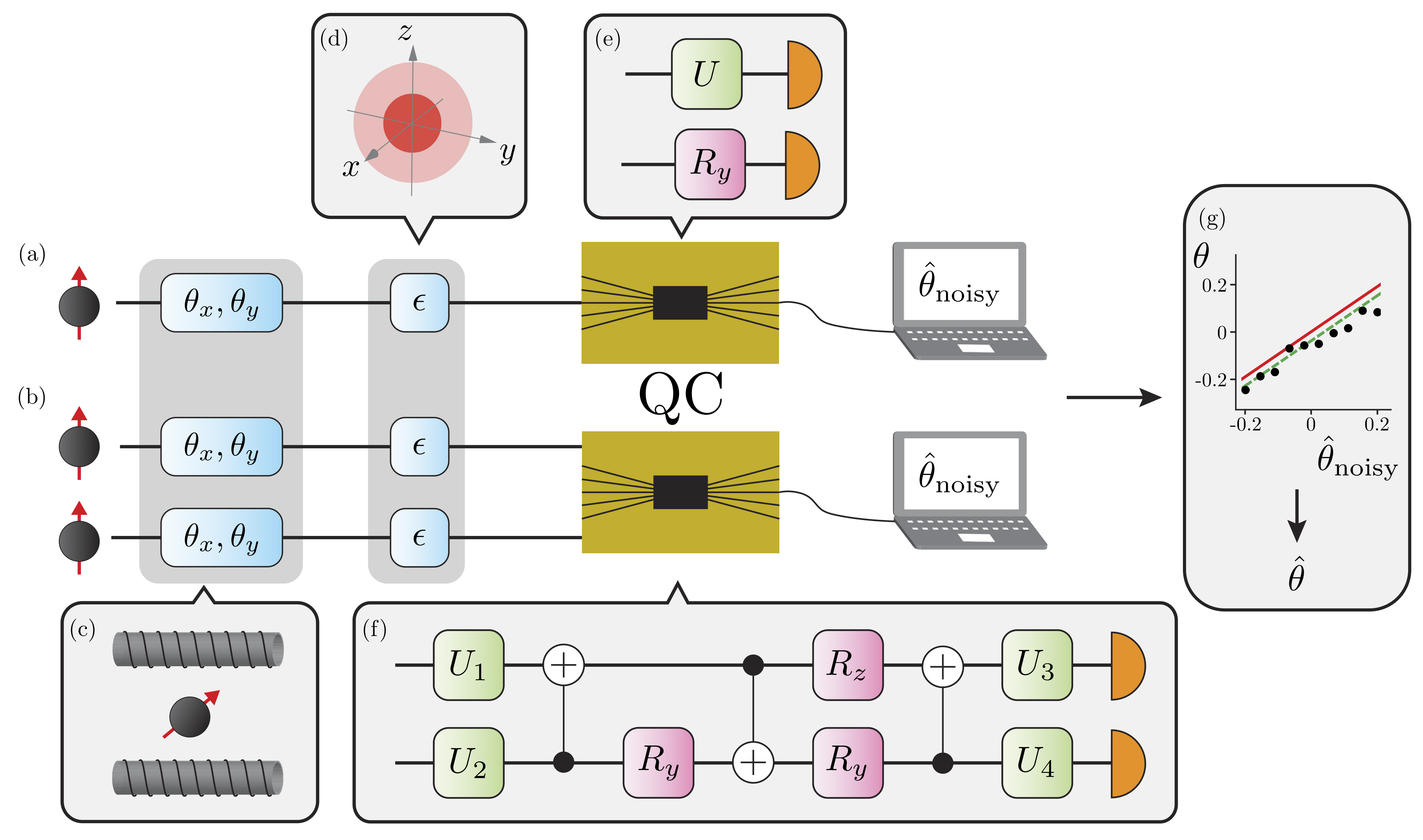}
\caption{\textbf{Experimental implementation of optimal collective measurements using quantum computers. }Probe states are sent to the quantum computers (QC) individually for the single-copy measurement (a) and in pairs for the two-copy measurement (b). The qubit probes experience rotations, $\theta_x$ and $\theta_y$, about the $x$ and $y$ axes of the Bloch sphere (c) before undergoing decoherence which has the effect of shrinking the Bloch vector (d). This rotation can be thought of as being caused by an external magnetic field which we wish to sense. The QCs then implement quantum circuits corresponding to the optimal single-copy (e) and two-copy (f) measurements. In (e) there are two optimal single-copy circuits, one for estimating $\theta_x$ and one for $\theta_y$. Finally, error mitigation is used to improve the accuracy of the estimated angle (g). We create a model (green line) for how the noisy estimate of $\theta$, $\hat{\theta}_\text{noisy}$ (black dots), is related to the true value (red line). The model is then used to correct $\hat{\theta}_\text{noisy}$ to produce the final estimate $\hat{\theta}$. Sample data from the \mbox{F-IBM QS1} device downsampled by a factor of 3 is shown in (g). Error bars are smaller than the markers.
}
\label{fig:schemetot}
\end{figure*}

Quantum-enhanced single parameter estimation is an established capability, with non-classical probe states achieving precisions beyond what can be reached by the equivalent classical resources in photonic~\cite{kacprowicz2010experimental,slussarenko2017unconditional,guo2020distributed}, trapped ion~\cite{mccormick2019quantum,leibfried2004toward}, superconducting~\cite{wang2019heisenberg} and atomic~\cite{muessel2014scalable,gross2010nonlinear} systems. This has paved the way for quantum enhancements in practical sensing applications, from gravitational wave detection~\cite{aasi2013enhanced} to biological imaging~\cite{casacio2021quantum}. For single-parameter estimation, entangled probe states are sufficient to reach the ultimate allowed precisions. However, for multi-parameter estimation, owing to the possible incompatibility of different observables, entangling resources are also required at the measurement stage. 
The ultimate attainable limits in quantum multi-parameter estimation are set by the Holevo Cram\'{e}r-Rao bound (Holevo bound)~\cite{holevo1973statistical,holevo2011probabilistic}. In most practical scenarios it is not feasible to reach the Holevo bound as this requires a collective measurement on infinitely many copies of the quantum state~\cite{kahn2009local, yamagata2013quantum, yang2019attaining,conlon2022gap} (see Methods M1 for a rigorous definition of collective measurements). Nevertheless, it is important to develop techniques which will enable the Holevo bound to be approached, given that multi-parameter estimation is fundamentally connected to the uncertainty principle~\cite{heisenberg1985anschaulichen} and has many physically motivated applications, including simultaneously estimating phase and phase diffusion~\cite{vidrighin2014joint,szczykulska2017reaching}, quantum super-resolution~\cite{vrehavcek2017multiparameter,chrostowski2017super}, estimating the components of a 3D field~\cite{baumgratz2016quantum,hou2020minimal} and tracking chemical processes~\cite{cimini2019quantum}. Furthermore, as we demonstrate,  collective measurements offer an avenue to quantum-enhanced sensing even in the presence of large amounts of decoherence, unlike the use of entangled probe states~\cite{dorner2009optimal,demkowicz2012elusive}.


To-date, collective measurements for quantum multi-parameter metrology have been demonstrated exclusively on optical systems~\cite{roccia2017entangling,parniak2018beating,hou2018deterministic,wu2019experimentally,yuan2020direct,wu2020minimizing}. Contemporary approaches to collective measurements on optical systems are limited in their scalability, i.e.\ it is difficult to generalise present approaches to measuring many copies of a quantum state simultaneously. The limited gate set available can also make it harder to implement an arbitrary optimal measurement. Indeed, the collective measurements demonstrated so far have all been restricted to measuring two copies of the quantum state and, while quantum enhancement has been observed, have all failed to reach the ultimate theoretical limits on separable measurements~\cite{nagaoka2005new,nagaoka2005generalization}.
Thus, there is a pressing need for a more versatile and scalable approach to implementing collective measurements.

\begin{figure*}[t]
\includegraphics[width=\textwidth]{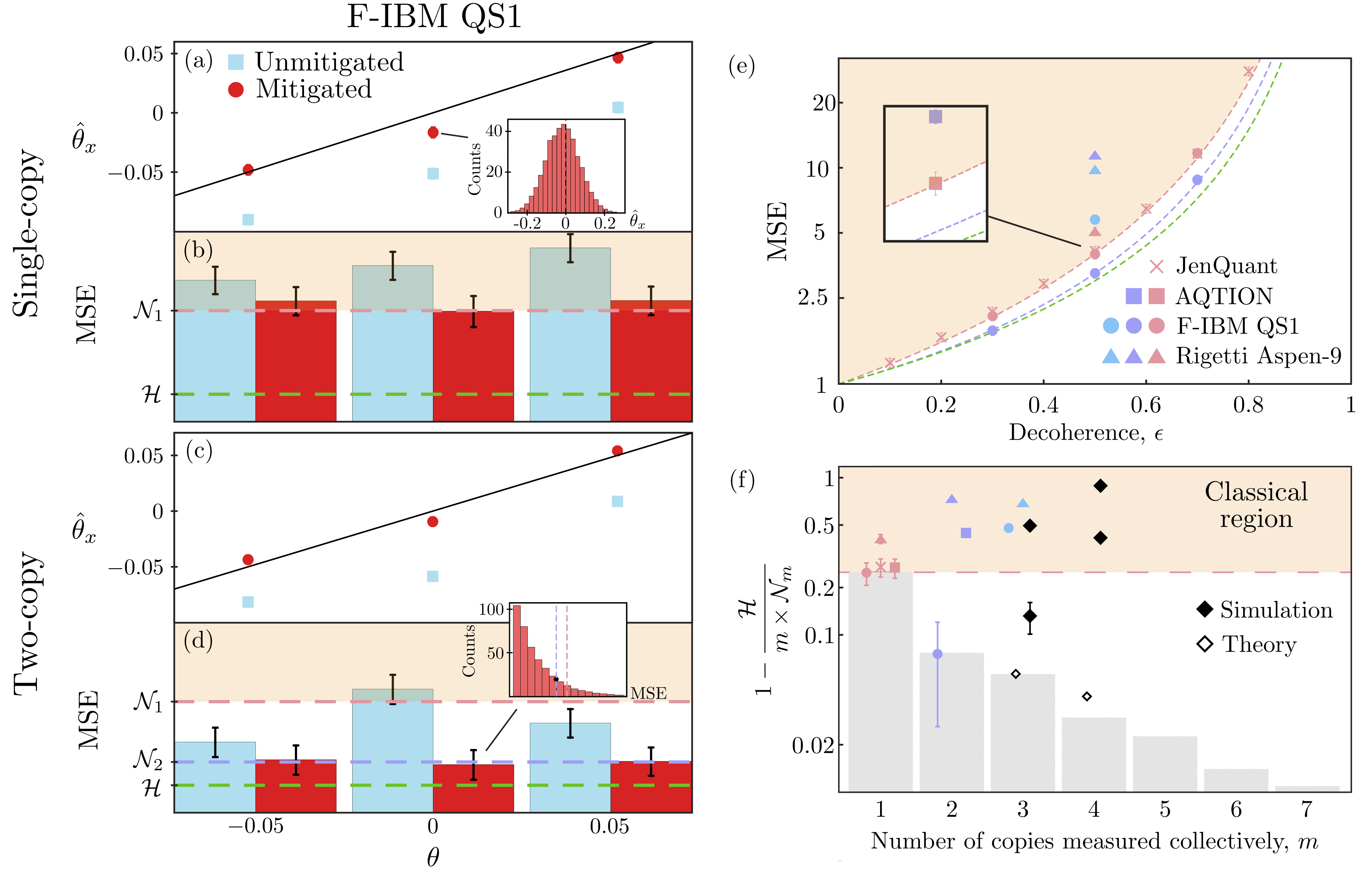}
\caption{\textbf{Surpassing single-copy limits through collective measurements.} In all figures \protect\markerone, \protect\markertwo\space and \protect\markerthree\space correspond to the single-copy Nagaoka, two-copy Nagaoka and Holevo bounds respectively. The orange-shaded region corresponds to the mean squared error (MSE) attainable with separable measurements. MSEs below \protect\markerthree\space are forbidden by quantum mechanics. Error bars are obtained using the bootstrapping technique~\cite{rice2006mathematical} and correspond to one standard deviation. All experimental points have error bars but some are smaller than the marker size. Each data point corresponds to the average of 400 individual experimental runs, each using 512 shots, as shown in the inset of (a) (see Methods M4). (a) and (c) show single-copy and two-copy estimates of $\theta_x$ respectively, both with and without error mitigation. Results for estimating $\theta_y$ are similar (Extended Data Fig.~1). (b) and (d) show the corresponding MSE. The distribution of MSE values follows the expected chi-squared distribution, shown in the inset of (d). The black circle in the inset corresponds to the mean MSE value. The results shown in (a)-(d) are for decoherence parameter $\epsilon=0.5$ and are obtained on the \mbox{F-IBM QS1} device. (e) Optimal single-, two- and three-copy measurements at different decoherence strengths, $\epsilon$. The pink, purple and blue markers correspond to experimental single-, two- and three-copy measurements respectively. For the superconducting devices all markers correspond to the precision after using error mitigation. The results of the AQTION trapped-ion processor for $\epsilon=0.5$ are shown in the inset for clarity. (f) Bars are one minus the ratio of the Holevo bound to the $m$-copy Nagaoka bound, for $m$ up to and including 7, calculated theoretically at $\epsilon=0.5$. Experimental points are one minus the ratio of the Holevo bound to the MSE obtained experimentally. \protect\showpgfdiamondnofill\space corresponds to the precision which our three- and four-copy projective measurements can obtain in theory. The upper and lower \protect\showpgfdiamond\space are simulations based on a depolarising noise model with gate error rates of $5\times10^{-3}$ and $1\times10^{-3}$ respectively. Legend is the same as in (e). }
\label{fig:emres}
\end{figure*}

In this work we design and implement theoretically optimal collective measurement circuits on superconducting and trapped-ion platforms. The ease with which these devices can be reprogrammed, the universal gate set available, and the number of modes across which entanglement can be generated, ensure that they avoid many of the issues that current optical systems suffer from. Using recently developed error mitigation techniques~\cite{czarnik2021error} we estimate qubit rotations about the axes of the Bloch sphere with a greater precision than what is allowed by separable measurements on individual qubits. This approach allows us to investigate several interesting physical phenomena: i) We demonstrate both optimal single-copy and two-copy collective measurements reaching the theoretical limits~\cite{nagaoka2005new,nagaoka2005generalization}. 
We also implement a three-copy collective measurement as a first step towards surpassing two-copy measurements. However, due to the circuit complexity, this measurement performs worse than single-copy measurements. ii) We investigate the connection between collective measurements and the uncertainty principle. Using two-copy collective measurements, we experimentally violate a metrological bound based on known, but restrictive uncertainty relations~\cite{lu2021incorporating}. iii) Finally, we compare the metrological performance of quantum processors from different platforms, providing an indication of how future quantum metrology networks may look.

%
\vspace*{-1em}
\subsection*{Theoretical Results}
In this work we implement theoretically optimal quantum circuits saturating the Nagaoka bound~\cite{nagaoka2005new,nagaoka2005generalization}, which sets an upper limit on the precision attainable with separable measurements. We consider the probe $\ket{\psi}=\ket{0}$, which experiences small rotations, $\theta_x$ and $\theta_y$, about the $x$ and $y$ axes of the Bloch sphere respectively before getting decohered (Fig.~1 (c) and (d)). For small rotations, the state becomes $\rho_{1}\approx(1-\epsilon)\ket{0}\bra{0}+\epsilon/2\mathbb{1}$, where $\epsilon$ is the decoherence strength. Such a noise model is relevant for quantum computing~\cite{vovrosh2021simple} and communication~\cite{bennett1999entanglement} among other applications. The Nagaoka bound is given by
\begin{equation}
\label{Nag12}
v_x+v_y \geq \mathcal{N}_{1}=\frac{4}{(1-\epsilon)^{2}}\;,
\end{equation} 
where $v_{x(y)}$ is the variance in the estimate of $\theta_{x(y)}$. This applies when the probe states are measured one by one (Fig.~1 (a)). We shall refer to measurements of this type as \textit{single-copy measurements}. The two-copy Nagaoka bound is
\begin{equation}
\label{Nag22}
v_x+v_y \geq \mathcal{N}_{2}=\frac{4-2\epsilon+\epsilon^2}{2(1-\epsilon)^2}\;,
\end{equation} 
which applies when we can perform a collective measurement on two copies of the probe, $\rho_{2}=\rho_{1}\otimes\rho_{1}$, which are entangled during the measurement (Fig.~1 (b)). These measurements are referred to as \textit{two-copy measurements}. Finally, allowing for collective measurements on infinitely many copies of the probe state, the Holevo bound is
\begin{equation}
\label{eq:hol}
v_x+v_y \geq \mathcal{H}=\lim_{m\to\infty}m\times\mathcal{N}_m=\frac{4-2\epsilon}{(1-\epsilon)^2}\;.
\end{equation} 
The hierarchy between the bounds is, $ \mathcal{H}\leq2\mathcal{N}_{2}\leq\mathcal{N}_{1}$, with equality only for $\epsilon=0$ or $1$. Detail on the computation of the different bounds is given in Supplementary Note 1. 

The Nagaoka bounds, Eqs.~\eqref{Nag12} and \eqref{Nag22}, can be saturated by positive operator valued measures (POVMs) in 2- and 4-dimensional Hilbert spaces respectively (detailed in Supplementary Note 2). 
For single-copy measurements, it is possible to measure $\theta_x$ and $\theta_y$ separately, with two different POVMs, each using half of the total probe states without any loss in precision (Fig.~1 (e)). For the two-copy measurement this is not possible; both parameters have to be estimated simultaneously to take advantage of the collective measurement. In order to implement the optimal POVMs, we find a unitary matrix which diagonalises each POVM in the computational basis. Using standard techniques from quantum computing, we then convert these unitary matrices to quantum circuits~\cite{vatan2004optimal}, which can be executed experimentally (Fig.~1 (e) and (f)). We present three- and four-copy POVMs, and the corresponding quantum circuits, which theoretically surpass the two- and three-copy Nagaoka bounds respectively, in Supplementary Notes 3 and 4.
%

We also investigate the asymptotic attainability of the Holevo bound, examining how closely measurements on a finite number of copies of the probe state can approach it. In Fig.~2 (f), we compute the Nagaoka bound for performing collective measurements on up to 7 copies of the probe state simultaneously, corresponding to a 128-dimensional Hilbert space~\cite{conlon2021efficient}.


\subsection*{Experimental Results}
In what follows we will describe the results of experiments conducted on multiple quantum platforms. The superconducting processors used were the Fraunhofer IBM Q System One (\mbox{F-IBM QS1}) processor, 11 cloud-accessible IBM Q processors and the Rigetti Aspen-9 processor. The trapped-ion (AQTION) processor is described in Ref.~\cite{pogorelov2021compact} and the Jena quantum photonic processor (JenQuant) is described in Methods M2. We implement the circuits corresponding to the optimal POVMs, shown in Fig.~1 (e) and (f), on the superconducting and trapped-ion processors. Additionally, we implement the single-copy measurements on JenQuant. The specific circuit parameters are provided in Supplementary Note 4. The outcomes of each run of a circuit are input to an estimator function to return the estimated values $\hat{\theta}_x$ and $\hat{\theta}_y$. This allows the mean squared error (MSE) to be determined.

\subsubsection*{Error Mitigation for Quantum Metrology}
Our first experiment investigates one possible application of error mitigation to quantum metrology. The details of the error mitigation used are found in Methods M4, but it is essentially a calibration process based on known angles as shown in Fig.~1 (g). For this experiment, conducted on the \mbox{F-IBM QS1} processor, the decoherence parameter is fixed at $\epsilon=0.5$ and we estimate a range of $\theta$ values. This verifies the unbiasedness of the estimator after error mitigation. Fig.~2 (a) and (c) show the average estimate of $\theta_x$, both before and after error mitigation, with single- and two-copy measurements respectively. The improvement offered by error mitigation, evident in these figures, is quantified by the MSE in \mbox{Fig.~2 (b)} and (d). Error mitigation cannot reduce the MSE below what is theoretically allowed by the Nagaoka bound, but it does enable both the single-copy and two-copy measurements to reach the corresponding Nagaoka bounds. Crucially, Fig.~2 (d) shows the advantage of the two-copy measurement, achieving a precision beyond what is classically possible over the range of $\theta$ considered. This is the first experimental demonstration of a measurement saturating the two-copy Nagaoka bound. Averaged over the entire range of $\theta$, the two-copy measurements show a MSE \mbox{$19\pm4\text{ }\%$} below the theoretical single-copy measurement limit, which is only \mbox{$6 \pm4\text{ }\%$} larger than the Holevo bound. In contrast, when restricted to single-copy measurements, the MSE is guaranteed to be at least \mbox{$33\text{ }\%$} larger than the Holevo bound. The ability to measure a range of angles is important for practical applications of quantum-enhanced metrology.

\subsubsection*{Optimal Single-, Two- and Three-Copy Measurements}
We next fix the rotations to $\theta_x=\theta_y=0$ and demonstrate a quantum enhancement over a range of $\epsilon$ values. Fig.~2 (e) shows the (scaled) MSE attained on different platforms. Using the \mbox{F-IBM QS1} device we are able to demonstrate a clear quantum enhancement across a range of $\epsilon$ values. The two-copy measurement on the \mbox{F-IBM QS1} device shows a maximum advantage over the theoretical single-copy limit of \mbox{$21\pm4\text{ }\%$}. In contrast, the Rigetti Aspen-9 superconducting device does not approach the theoretical limits for any of the measurements, likely due to the higher gate and readout error rates. Notably, both JenQuant and the AQTION processor are able to reach the theoretical single-copy measurement limits without any error mitigation. The AQTION processor does not however, reach the theoretical two-copy limits. The demonstration of quantum-enhanced sensing with highly mixed states showcases that collective measurements may provide metrological gain in real-world sensing applications where decoherence is unavoidable.

In Fig.~2 (e) and (f), we show the MSE of our three-copy measurement when implemented on the Rigetti Aspen-9 and \mbox{F-IBM QS1} processors. In Supplementary Note 6 we present further three-copy results for these and several other devices, all of which failed to reach the theoretical limit and display properties of a bad estimator. These experimental results are in qualitative agreement with simulations of three- and four-copy measurements based on the noise level expected for near-future quantum processors, also shown in Fig.~2 (f). From Fig.~2 (f), it is evident that for the problem we have considered, the benefit of three-copy measurements over two-copy measurements is marginal. This raises the question of whether measurements on many copies of a quantum state simultaneously are practically useful. In Supplementary Note 7 we present a similar problem, based on an amplitude damping noise model, where there is a sizeable gap between the two-copy Nagaoka and Holevo bounds, suggesting that collective measurements on many copies may be useful. With continually decreasing error rates, superconducting and trapped-ion devices may bridge this gap and approach the Holevo bound ever more closely. However, as the data from Fig.~2 (f) shows, there is a pertinent trade-off between what is gained by measuring more copies of the quantum state and what is lost by the increased experimental complexity. 

 \begin{figure}[t]
\includegraphics[width=0.4\textwidth]{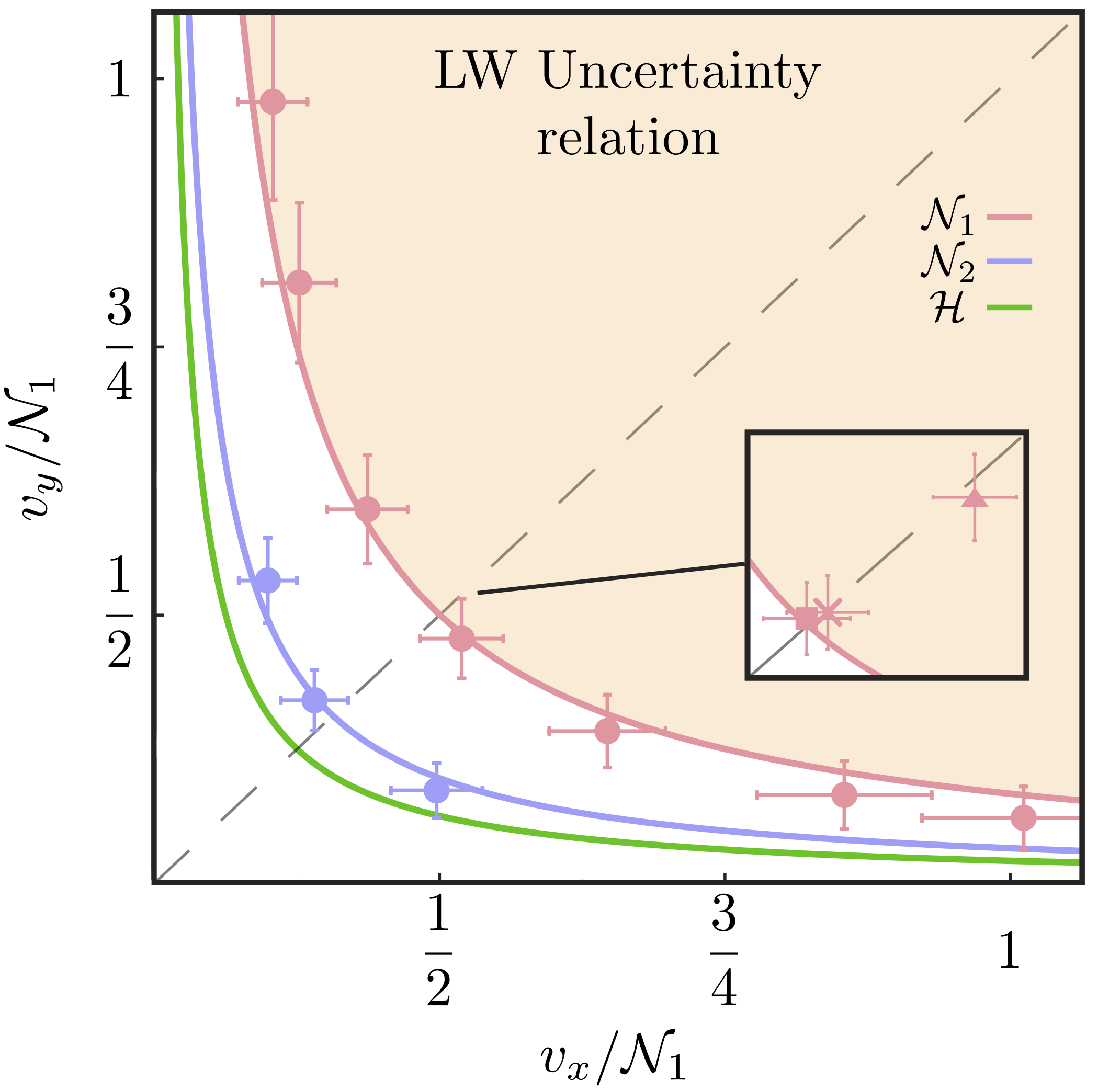}
\caption{\textbf{Collective measurements violating the LW uncertainty relation.} The shaded region shows the measurement variances allowed by the LW uncertainty relation~\cite{lu2021incorporating}. All experimental points correspond to a decoherence parameter of $\epsilon=0.5$. 
The dashed grey line shows where the variance in estimating both parameters are equal. 
The purple and green lines are obtained by calculating the two-copy Nagaoka bound and Holevo bound with different weights attached to each parameter. Solid lines are used for bounds on the allowed values ($v_x,v_y$) (as opposed to the sum of variances as in Fig.~2). The data in the main figure corresponds to the \mbox{F-IBM QS1} device. Data from the other processors is shown in the inset. The legend is the same as Fig.~2 (e). Each data point corresponds to the average of 400 individual experimental runs, each using 512 shots and error bars correspond to one standard deviation obtained by bootstrapping.}
\label{figucprincp}
\end{figure}

\begin{figure*}[t]
\includegraphics[width=\textwidth]{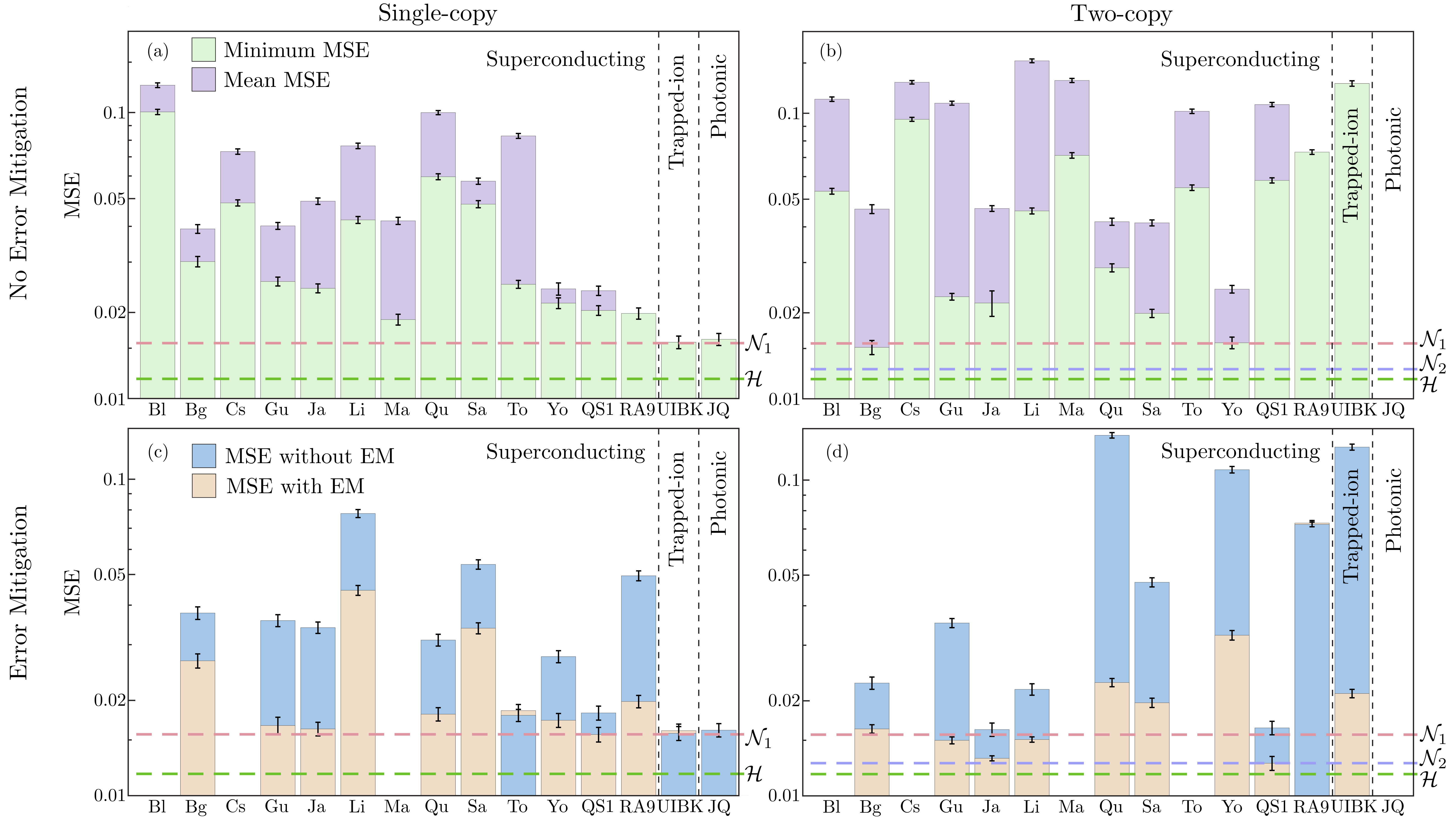}
\caption{\textbf{Comparing optimal measurement circuits on different quantum processors.} (a) and (b) show the mean MSE and minimum MSE across all qubits with different quantum processors for the single-copy and two-copy measurements respectively. No error mitigation is used in these figures. Each MSE is averaged over 600 experimental runs, corresponding to 5 different angles, each using 512 shots. (c) and (d) show the MSE with and without error mitigation (EM) for the single-copy and two-copy measurements respectively. In all but four cases error mitigation is beneficial. The data in (c) and (d) corresponds to the average of 400 individual experimental runs, each using 512 shots. For all figures, error bars correspond to one standard deviation obtained by bootstrapping. The different IBM quantum computers tested are Belem (Bl), Bogota (Bg), Casablanca (Cs), Guadalupe (Gu), Jakarta (Ja), Lima (Li), Manhattan (Ma), Quito (Qu), Santiago (Sa), Toronto (To), Yorktown (Yo) and the \mbox{F-IBM QS1} device (QS1). Also shown is the Rigetti Aspen-9 superconducting device (RA9), JenQuant (JQ) and the AQTION trapped-ion device (UBIK). For the Rigetti Aspen-9 device, only one qubit / pair of qubits was tested, hence the mean MSE and minimum MSE are equal. For the AQTION device the mean MSE and minimum MSE are equal as only one ion / pair of ions was loaded into the trap. Empty spaces correspond to processors where a particular experiment could not be carried out.
}
\label{fig:benchmark}
\end{figure*}

 \subsubsection*{Collective Measurements and the Uncertainty Principle}
The uncertainty principle is one of the most fundamental features of quantum mechanics~\cite{heisenberg1985anschaulichen}. Recently, it has been observed that the original formulations of the uncertainty principle fail to hold in certain scenarios~\cite{erhart2012experimental,rozema2012violation}, leading to the introduction of \textquote{universally valid} uncertainty relations (UVUR) for operators~\cite{ozawa2003universally,ozawa2004uncertainty,branciard2013error}. In spite of the name, UVUR assume that measurements are carried out on single copies of the quantum state. This appears to be a natural assumption when considering how the measurement of one quantity disturbs any subsequent measurement of a second quantity. However, the same is not true when considering the precision with which two quantities can be jointly measured. Given this restriction, one might expect that UVUR can be violated through collective measurements.

Recently Lu and Wang extended the UVUR to quantum multi-parameter estimation~\cite{lu2021incorporating}, deriving a metrological bound on how well two parameters can be simultaneously estimated. We shall denote this as the LW uncertainty relation. %
%
%
For our problem this bound can be calculated as (see Supplementary Note 8)
\begin{equation}
 \label{eq:LW}
 \frac{1}{v_x}+\frac{1}{v_y}\leq(1-\epsilon)^2\;,
 \end{equation}
 which is saturated when $v_x=v_y=2/(1-\epsilon)^2$. The variances allowed by Eq.~\eqref{eq:LW} coincide with our single-copy measurement variances. Indeed, our single-copy measurement variances, shown in pink in Fig.~3, verify the validity of UVUR in this scenario. However, our two-copy measurements implemented on the \mbox{F-IBM QS1} processor were able to experimentally violate the LW uncertainty relation by more than 3 standard deviations as shown in purple in Fig.~3. The POVMs which give rise to the unbalanced variances are presented in Supplementary Note 9. To the best of our knowledge, this is the first time it has been observed that UVUR can be surpassed, either theoretically or experimentally. These results have significance for the manner in which the uncertainty principle is interpreted and suggest that tighter uncertainty relations are required when allowing for any measurement type. In Supplementary Note 10 we relate the violation of the LW uncertainty relation to the more common error-disturbance operator uncertainty relations.

\subsubsection*{Cross-platform Comparison}

Our final experiment compares the performance of different platforms for estimating qubit rotations. This provides an indication of what resources may be used in a future quantum metrology network.
For superconducting devices, we first perform simultaneous qubit rotation estimation using all non-neighbouring (pairs of) qubits, to minimise cross-talk between qubits. The mean MSE and minimum MSE across all qubits is shown in Fig.~4 (a) and (b) for each device tested. Each MSE is averaged over estimating 5 angles in the range $\theta=-0.01$ to $0.01$, repeated 120 times for each angle. For the trapped-ion and photonic devices only one photon, ion or pair of ions was used, hence only the mean MSE is shown. We then repeat the experiment using only the best performing qubit(s), now applying error mitigation as shown in Fig.~4 (c) and (d). The benefits of error mitigation are most pronounced for the \mbox{F-IBM QS1} processor as we had unrestricted access to this device. Having restricted access to a device means each experiment takes longer, hence the model for the device provided by error mitigation is likely to be less accurate by the end of the experiment. 

\subsection*{Discussion}
Superconducting and trapped-ion devices are natural platforms for attaining the maximal advantage of quantum metrology and quantum information tasks through collective measurements. By implementing collective measurements on pairs of quantum states, we have been able to perform quantum multi-parameter estimation with a precision that cannot be reached classically using the same resources. There are many scenarios where this work may prove beneficial, particularly when there is an intrinsic restriction on resources. One can envision an optical system connected to a quantum processor through optical-to-microwave converters~\cite{higginbotham2018harnessing}. With only a limited number of qubits, such a device could greatly enhance biomedical imaging or quantum communications, meaning these advantages may be leveraged with near-future technology. Furthermore, collective measurements can be beneficial for quantum tomography~\cite{massar2005optimal}, entanglement distillation for quantum communication~\cite{bennett1996purification} and quantum illumination~\cite{zhuang2017optimum}.


This work opens up a number of avenues for future investigation: a natural extension to using error mitigation for quantum metrology is error correction~\cite{dur2014improved}. With the aid of the techniques presented here, it may be possible to demonstrate multi-parameter metrology which fully utilises quantum resources; benefiting from both entangled probe states and collective measurements. 
By simplifying our three-copy measurement circuit, the theoretical limits may be approachable with the present generation of quantum processors. It would also be pertinent to study further how gate error rates and circuit complexity need to scale to successfully implement many-copy collective measurements. Investigating further the connection between collective measurements and the uncertainty principle may reveal important aspects of fundamental physics and could lead to the development of tighter uncertainty relations which hold true for any measurement type. Finally, the ideal extension of our work is to demonstrate optimal collective measurements in a practical setting. We anticipate that our work brings this closer.

\section*{Acknowledgements}
We acknowledge the use of IBM Quantum services and the Rigetti  Aspen-9 processor for this work. The views expressed are those of the authors, and do not reflect the official policy or position of IBM, the IBM Quantum team or Rigetti.

The authors would like to acknowledge the support of Amazon Web Services by making available the Rigetti Aspen-9 device using the Amazon Braket service.

This research was funded by the Australian Research Council Centre of Excellence CE170100012, Laureate Fellowship FL150100019 and the Australian Government Research Training Program Scholarship.

This work has been supported by the Fraunhofer-Gesellschaft zur Förderung der angewandten Forschung e.V. We gratefully acknowledge the financial support by the German Federal Ministry of Education and Research via \textquote{2D Nanomaterialien für die Nanoskopie der Zukunft}, FKZ: 13XP5053A and the European Union, the European Social Funds and the Federal State of Thuringia under Grant ID 2021FGI0043.

This work was funded by the Deutsche Forschungsgemeinschaft (DFG, German Research Foundation) - Projektnummer 445275953. The authors acknowledge support by the German Space Agency DLR with funds provided by the Federal Ministry for Economic Affairs and Energy BMWi under grant number 50WM2165 (QUICK3).

DWB is supported by Australian Research Council Discovery Projects DP190102633 and DP210101367.

CDM, IP, and TM acknowledge funding from the EU H2020-FETFLAG-2018-03 under Grant Agreement no. 820495. We also acknowledge support by the Austrian Science Fund (FWF), through the SFB BeyondC (FWF Project No.\ F7109), and the IQI GmbH, as well as the Office of the Director of National Intelligence (ODNI), Intelligence Advanced Research Projects Activity (IARPA), via US ARO grant no. W911NF-16-1-0070 and W911NF-20-1-0007, and the US Air Force Office of Scientific Re-search (AFOSR) via IOE Grant No. FA9550-19-1-7044 LASCEM.

\section*{Author contributions}
PKL conceived the project. LC and SMA derived the theoretical results and designed the optimal quantum circuits. LC, DWB and SMA optimised the quantum circuits. TV ran the JenQuant experiment. CDM and IP ran the AQTION experiment. LC and FE ran the \mbox{F-IBM QS1} experiment. LC performed the data analysis. LC, SY and SMA derived the bounds in Fig.~3. LC wrote the manuscript with contributions from all authors.
\section*{Competing interests}
The authors declare no competing interests.

\bibliography{col_meas_bib}
\bibliographystyle{naturemag}

\section*{Methods}

\subsection{Collective measurements}
\label{method:CM}

Here we clarify our use of terminology regarding \textquote{entangling} and \textquote{collective} measurements. We stick to the definitions used in Refs.~\cite{roccia2017entangling,parniak2018beating,hou2018deterministic,wu2019experimentally,yuan2020direct,wu2020minimizing}, where a collective measurement is a measurement that acts on multiple copies of the quantum state simultaneously. An N-copy collective measurement thus simultaneously measures N copies of the same state, whereas a \textquote{single-copy} measurement, or \textquote{separable} measurement, measures the quantum states individually. The quantum states themselves may consist of an arbitrary number of possibly entangled modes. When we refer to \textquote{entangling} measurements we mean measurements capable of creating entanglement between \textit{multiple copies} of the quantum state, or alternatively, in an entangled multi-copy basis.

There are many similar concepts, which may be confused with our definition of a collective measurement. For example: In Ref.~\citelatex{marciniak2021optimal} a quantum state with 26 entangled modes (ions) was used. In our terminology, measuring the 26 ions simultaneously is a separable measurement, because only a single copy of the quantum state was used and consequently no entanglement between copies was possible. However, in principle the (0,2) and (1,2) schemes in Ref.~\citelatex{marciniak2021optimal} could be used for implementing collective measurements in the sense of our definition. Similarly, Ref.~\citelatex{bohnet2014reduced} refers to collective measurements as measurements of ensemble quantities of atoms, wholly unrelated to our terminology. In Ref.~\citelatex{jagannathan2022demonstration} multi-copy discrimination of two quantum states is demonstrated. However, this multi-copy discrimination uses separable measurements, the multi-copy part referring to the fact that multiple (separable) measurement outcomes are used in making a final decision. Finally, Refs.~\citelatex{toth2020activating,trenyi2022multicopy} examine multi-copy metrology. Again in this work, the term multi-copy carries a different meaning compared to our work, as only single parameter estimation was considered.

\subsection{Photonic experiment}
The Jena quantum photonic processor (JenQuant) is based on a single photon emitting colour centre in the two-dimensional material hexagonal boron nitride (hBN). The crystal defect introduces an effective two-level system into the bandgap that is excited optically. The emitter is fabricated by treating a multilayer hBN crystal with an oxygen plasma and subsequent rapid thermal annealing \citelatex{10.1021/acsphotonics.8b00127}. A suitable quantum emitter was then coupled to a hemispherical microcavity \citelatex{10.1021/acsphotonics.9b00314}. The resonator enhances the emission via the Purcell effect and suppresses noise to reduce the multi-photon probability below 0.6\% at room temperature \citelatex{PhysRevResearch.3.013296}. The spectrum is tunable by adjusting the resonator length within the free space emission linewidth of 5.76 nm (FWHM) around 565 nm and has a linewidth of 0.2 nm \citelatex{10.1021/acsphotonics.9b00314}. 

We encode the logical qubits in the polarisation of the photons and choose $\ket{\text{H/V}}$ as the computational basis states $\ket{0/1}$. The input states $\ket{0}$, $\ket{1}$, and $\ket{\psi_\theta}$ are set by motorized polarisation optics (a half-wave plate (HWP), polariser, and a quarter-wave plate (QWP)). The polariser ensures a high polarisation extinction ratio of $>10^5$:1. The single-copy POVMs are implemented by the combination of motorized QWP, HWP, QWP, which can perform any arbitrary unitary rotation. In Supplementary Note 4 we show the decomposition of the optimal single-copy POVMs into wave plate rotations. Finally, a polarising beam splitter projects onto the computational basis and the photons are detected by two single photon detectors in both arms. JenQuant is thereby a fully universal single qubit quantum computer. Performing multi-qubit operations requires an entangling gate such as CNOT which would require indistinguishable single photons. This in turn can be achieved by a narrower resonator linewidth $<124$ MHz to reach a Hong-Ou-Mandel contrast $>90$\% \citelatex{10.1021/acsphotonics.9b00314}. Note that JenQuant does not require any error mitigation, partly due to the long-term stability of the system.

\subsection{Superconducting experiments}
The \mbox{F-IBM QS1} device used is based in Ehningen. It uses an IBM Quantum Falcon processor and has 27 qubits. As with all IBM Quantum devices, the qubits are transmons. The frequency of the transmons are around 5 GHz~\citelatex{jurcevic2021demonstration}.



\subsection{Error mitigation}
\label{meth:em}
Before running each experiment for estimating the unknown angles $\theta_x$ and $\theta_y$, we implement Clifford data regression error mitigation~\cite{czarnik2021error}. This involves constructing a model for how a noisy expectation value predicted by a quantum processor is related to the true expectation value. In general complex models can be used, however, for quantum metrology, it is essential that the chosen model does not bias the estimator. We are therefore required to use a simple model of the form $\hat{\theta}_{x(y)}=\hat{\theta}_{\text{noisy, }x(y)}+c_{x(y)}$, where $\hat{\theta}_{\text{noisy, }x(y)}$ is the unmitigated $\theta_{x(y)}$ value predicted by the quantum processor and $c_{x(y)}$ is a constant. Detail on other possible models which were considered, but found to bias the estimator, is provided in Supplementary Note 5. We use 30 known $\theta$ values in the range $\theta\in[-0.2,0.2]$ rad to determine a value for the model $c_{x(y)}$. 
 An example of the model fitting is shown in Fig.~1 (g) for the \mbox{F-IBM QS1} quantum processor. This model is then used to estimate some unknown angle $\theta=\theta_x=\theta_y$. Unless otherwise specified in the main text, the model is recalibrated after every 40 predictions of the unknown angle and the process is repeated to estimate each unknown angle 400 times. Our figure of merit is taken to be the average MSE over all 400 runs.
\begin{equation}
\text{MSE}=\frac{1}{400}\sum_{i=1}^{400}((\theta_x-\hat{\theta}_{x,i})^2+(\theta_y-\hat{\theta}_{y,i})^2)\;,
\end{equation}
where $\hat{\theta}_{x(y),i}$ is the $i$th estimate of $\theta_{x(y)}$. To obtain each of the 400 estimates we average the results of 512 repetitions of the experiment for each of the single-copy circuits and for the two-copy circuit. For the three-copy circuit we average the results of 341 repetitions of the experiment to ensure equal resources are used in each experiment.

For the two-copy measurements in Fig.~3 with $v_x\neq v_y$, a slightly different error mitigation process was used. At the time this particular data was being taken it was not possible to recalibrate in between estimating the unknown angle. Hence, the calibration step was only performed once, immediately before estimating the unknown angle. To increase the utility of the error mitigation in this case, we used 30 known angles in the range $\theta\in[-0.05,0.05]$ rad. 

\section*{Data availability}
All data and codes are available at the following Github repository: https://github.com/LorcanConlon/Approaching-optimal-entangling-collective-measurements.
\section*{Code availability}
All data and codes are available at the following Github repository: https://github.com/LorcanConlon/Approaching-optimal-entangling-collective-measurements.

\section*{Extended Data}
\begin{figure*}[t]
\includegraphics[width=0.85\textwidth]{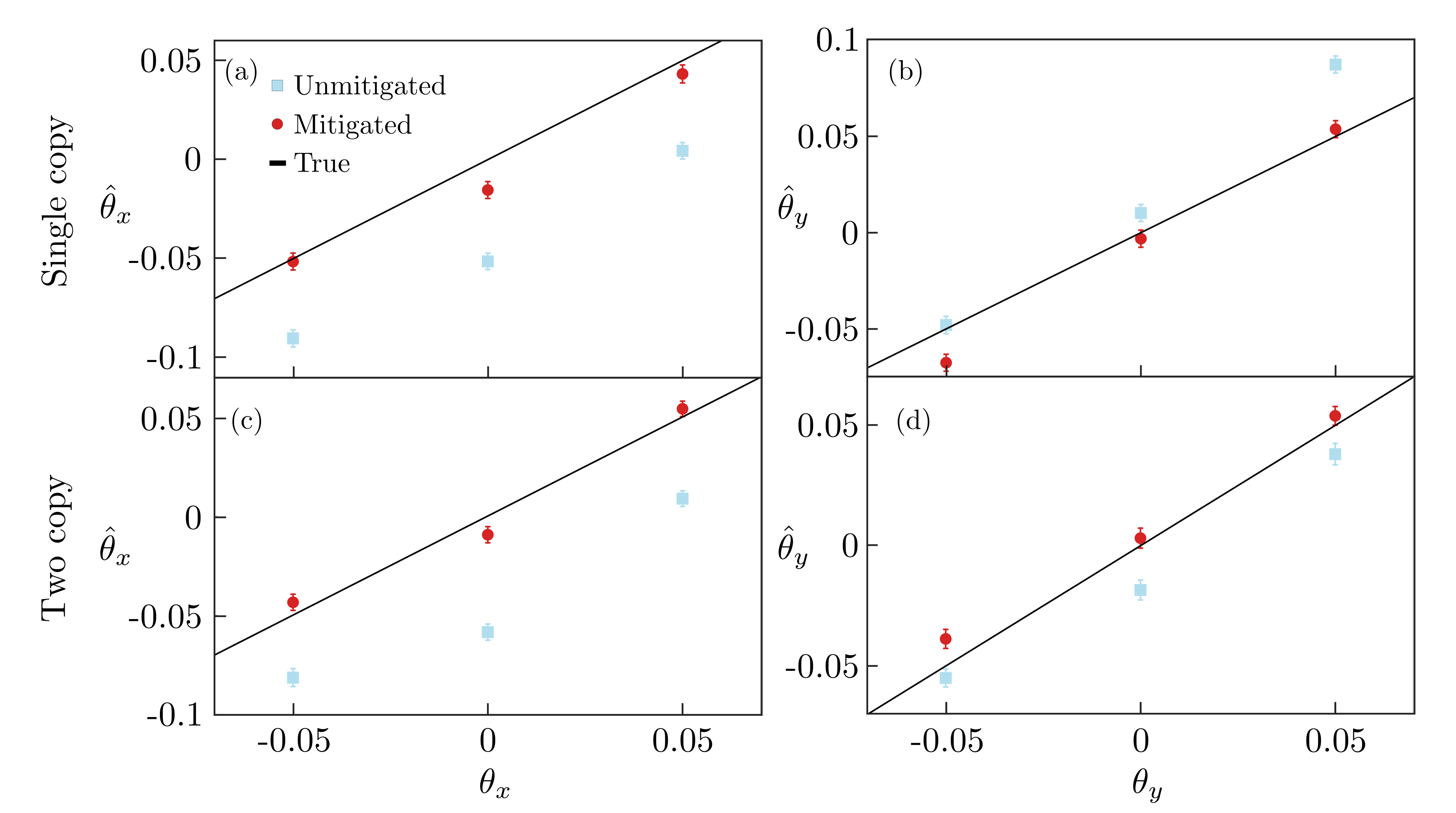}
\caption{\textbf{Effect of error mitigation on estimation performance.} Figs (a) to (d) show the estimated values of $\theta$, averaged over all 400 runs, before (blue squares) and after (red circles) applying error mitigation. Figs (a) and (b) ((c) and (d)) correspond to estimating $\theta_x$ and $\theta_y$ respectively with the optimal single(two)-copy measurement. Error bars are obtained using the bootstrapping technique~\cite{rice2006mathematical} and correspond to one standard deviation. All results shown are for decoherence parameter $\epsilon=0.5$ and are obtained on the \mbox{F-IBM QS1} device.}
\label{fig:emED}
\end{figure*}

\bibliographylatex{col_meas_bib_2}
\bibliographystylelatex{naturemag}

%


\end{document}


\title{Supplementary information : Approaching optimal entangling collective measurements on quantum computing platforms}

\author{Lorc{\'a}n O. Conlon}
\email{lorcanconlon@gmail.com}
\affiliation{\ANU}
\author{Tobias Vogl}
\affiliation{\Jena}
\affiliation{\Camb}
\author{Christian D. Marciniak}
\affiliation{\Inns}
\author{Ivan Pogorelov}
\affiliation{\Inns}
\author{Simon K. Yung}
\affiliation{\ANU}
\author{Falk Eilenberger}
\affiliation{\Jena}
\affiliation{\Fraun}
\affiliation{\MP}
\author{Dominic W. Berry}
\affiliation{\Sydney}
\author{Fabiana S. Santana }
\affiliation{\AWS}
\author{Rainer Blatt}
\affiliation{\Inns}
\affiliation{\Innstwo}
\author{Thomas Monz}
\affiliation{\Inns}
\affiliation{\Innsthree}
\author{Ping Koy Lam}
\email{ping.lam@anu.edu.au}
\affiliation{\ANU}
\affiliation{\NTU}
\affiliation{\Astar}
\author{Syed M. Assad}
\email{cqtsma@gmail.com}
\affiliation{\ANU}
\affiliation{\NTU}


\tableofcontents
\newpage
\subsection{Computation of the Holevo and Nagaoka bounds}
\label{SN1}
In this Supplementary Note we describe the computation of the single-copy Nagaoka, two-copy Nagaoka and Holevo bounds. For the qubit rotation estimation problem discussed in this work we consider the state $\ket{0}$ subject to rotations $\theta_x$ and $\theta_y$ about the $x$ and $y$ axes of the Bloch sphere respectively. The rotation operators are given by
 
\begin{equation}
R_{x}(\theta)=\begin{pmatrix}
\text{cos}(\frac{\theta}{2})& -\mathrm{i}\text{ sin}(\frac{\theta}{2}) \\
-\mathrm{i}\text{ sin}(\frac{\theta}{2}) & \text{cos}(\frac{\theta}{2})
\end{pmatrix}\qquad\text{and}\qquad
R_{y}(\theta)=\begin{pmatrix}
\text{cos}(\frac{\theta}{2})& -\text{sin}(\frac{\theta}{2}) \\
\text{sin}(\frac{\theta}{2}) & \text{cos}(\frac{\theta}{2})
\end{pmatrix}\;.
\end{equation} 
After the rotations, the probe is subject to the decoherence channel. Assuming the rotations are small, the density matrix and its derivatives are given by
\begin{equation}
\label{singlecopychar}
\rho=\begin{pmatrix}
1-\frac{\epsilon}{2}&0\\
0&\frac{\epsilon}{2}
\end{pmatrix}
\qquad\text{and}\qquad
\partial_{x} \rho =\begin{pmatrix}
0&\frac{\mathrm{i}}{2}(1-\epsilon)\\
\frac{-\mathrm{i}}{2}(1-\epsilon)&0
\end{pmatrix}
\qquad\text{and}\qquad
\partial_{y} \rho =\begin{pmatrix}
0&\frac{(1-\epsilon)}{2}\\
\frac{(1-\epsilon)}{2}&0
\end{pmatrix}\;,
\end{equation}
where $\epsilon$ is the decoherence strength and $\partial_i$ represents the partial derivative with respect to the parameter $\theta_i$. The parameters $\theta_x$ and $\theta_y$ can be estimated using a positive operator valued measure (POVM). A POVM is described by a set of non-negative operators, $\{\Pi_{l}\}$, which sum to the identity, $\sum_{l}\Pi_l=\mathbb{I}$. Each measurement outcome occurs with a certain state-dependent probability, $p_l=\text{tr}[\rho_{\theta}\Pi_l]$. The series of measurement results are used to construct unbiased estimators, $\hat{\theta}_{x}$ and $\hat{\theta}_{y}$, for the parameters of interest, $\theta_{x}$ and $\theta_{y}$. Our goal is to minimise the mean squared error (MSE) between the true $\theta$ and the estimated value $\hat{\theta}$. The MSE matrix is given by
\begin{equation}
[V(\hat{\theta})]_{jk}=\sum_{l}(\hat{\theta}_{j}(l)-\theta_j)(\hat{\theta}_{k}(l)-\theta_k)p_{l}\;,
\end{equation}
where the sum is over all possible measurement outcomes. The Holevo and Nagaoka bounds are lower bounds for the trace of the MSE matrix when using collective and separable measurements respectively.

The Holevo bound is obtained by solving the following non-trivial minimisation problem~\cite{holevo1973statistical, holevo2011probabilistic}
\begin{align}
\label{eq_hol2}
\text{Tr}[V]\geq\mathcal{H} \coloneqq  \min_{X} \text{Tr}[Z_\theta[X]] +\text{TrAbs}[\Im Z_\theta[X]]\;,
\end{align}
where $Z_\theta[X]_{jk} \coloneqq \text{Tr} [\rho X_j X_k]$, $\text{Im}Z_{\theta}[X]$ represents the imaginary part of $Z_{\theta}[X]$, and $\text{TrAbs}[\text{Im}Z_{\theta}[X]]$ is the sum of the absolute values of the eigenvalues of the matrix $\text{Im}Z_{\theta}[X]$. The minimisation is over the Hermitian matrices $X$, subject to the unbiased conditions
\begin{equation}
\begin{split}
\label{eq:unbiased}
\text{Tr}[\rho X_i]&=0\\
\text{Tr}[\partial_i \rho X_j]&=\delta_{ij}\;.
\end{split}
\end{equation}
The Holevo bound applies when we allow for collective measurements on infinitely many copies of the probe state. When restricted to separable measurements, the Nagaoka bound provides an upper limit on the attainable precision. For estimating two parameters, the Nagaoka bound is given by~\cite{nagaoka2005new,nagaoka2005generalization}
\begin{equation}
\label{eq_hol2N}
  \text{Tr}[V]\geq \mathcal{N} \coloneqq  \min_{X} \text{Tr}[ Z_\theta[X]]
  +\text{TrAbs}[ \rho [X_1,X_2]]\;,
\end{equation}
where $[A,B]=AB-BA$ and the Hermitian matrices $X$ are subject to the same unbiased conditions as before. 

The following Hermitian matrices simultaneously solve the optimisation problem for the Holevo bound and the single-copy Nagaoka bound
\begin{equation}
X_x=\begin{pmatrix}
0&\frac{\mathrm{i}}{1-\epsilon}\\
\frac{-\mathrm{i}}{1-\epsilon}&0
\end{pmatrix}
\qquad\text{and}\qquad
X_y=\begin{pmatrix}
0&\frac{1}{1-\epsilon}\\
\frac{1}{1-\epsilon}&0
\end{pmatrix}\;.
\end{equation}
By substituting into Eq.~\eqref{eq_hol2} it is easily verified that these matrices give a Holevo bound of
\begin{align}
\label{eq:hol}
v_x+v_y \geq \mathcal{H}=\frac{4-2\epsilon}{(1-\epsilon)^2}\;,
\end{align}
where $v_{x(y)}$ is the variance in the estimate of $\theta_{x(y)}$. Similarly, by substituting into Eq.~\eqref{eq_hol2N}, the single-copy Nagaoka bound is given by
\begin{equation}
\label{Nag12}
v_x+v_y \geq \mathcal{N}_{1}=\frac{4}{(1-\epsilon)^{2}}\;.
\end{equation}

For computing the two-copy Nagaoka bound it is no longer sufficient to consider the state and derivatives in Eq.~\eqref{singlecopychar}. To find the attainable precision when performing collective measurements on two copies of the quantum state simultaneously, we make the transformation 
\begin{equation}
\rho\rightarrow\rho\otimes\rho, \qquad \partial_{x} \rho\rightarrow\partial_{x} \rho\otimes\rho+\rho\otimes\partial_{x} \rho\qquad\text{ and }\qquad \partial_{y} \rho\rightarrow\partial_{y} \rho\otimes\rho+\rho\otimes\partial_{y} \rho\;.
\end{equation}
The Hermitian matrices which solve the two-copy Nagaoka bound are given by
\begin{equation}
X_x=\begin{pmatrix}
0&\frac{\mathrm{i}}{2(1-\epsilon)}&\frac{\mathrm{i}}{2(1-\epsilon)}&\frac{1-\mathrm{i}}{2(1-\epsilon)}\\
\frac{-\mathrm{i}}{2(1-\epsilon)}&0&0&\frac{\mathrm{i}}{2(1-\epsilon)}\\
\frac{-\mathrm{i}}{2(1-\epsilon)}&0&0&\frac{\mathrm{i}}{2(1-\epsilon)}\\
\frac{1+\mathrm{i}}{2(1-\epsilon)}&\frac{-\mathrm{i}}{2(1-\epsilon)}&\frac{-\mathrm{i}}{2(1-\epsilon)}&0
\end{pmatrix}
\qquad\text{and}\qquad
X_y=\begin{pmatrix}
0&\frac{1}{2(1-\epsilon)}&\frac{1}{2(1-\epsilon)}&\frac{1+\mathrm{i}}{2(1-\epsilon)}\\
\frac{1}{2(1-\epsilon)}&0&0&\frac{1}{2(1-\epsilon)}\\
\frac{1}{2(1-\epsilon)}&0&0&\frac{1}{2(1-\epsilon)}\\
\frac{1-\mathrm{i}}{2(1-\epsilon)}&\frac{1}{2(1-\epsilon)}&\frac{1}{2(1-\epsilon)}&0
\end{pmatrix}\;.
\end{equation}
We note that these solutions are not unique and multiple optimal solutions were found for the two-copy Nagaoka bound. This solution gives the two-copy Nagaoka bound as
\begin{equation}
\label{Nag22}
v_x+v_y \geq \mathcal{N}_{2}=\frac{4-2\epsilon+\epsilon^2}{2(1-\epsilon)^2}\;.
\end{equation}
As both the Holevo and Nagaoka bounds can be computed efficiently~\cite{albarelli2019evaluating,conlon2021efficient} we can be sure our solutions are correct.

\subsection{POVMs saturating the Nagaoka bounds}
\label{SN2}
In this Supplementary Note we present measurement strategies, i.e. POVMs and estimator functions, which saturate the single- and two-copy Nagaoka bounds. A POVM which saturates the single-copy Nagaoka bound is given by
\begin{equation}
\begin{split}
\label{POVM1qDC}
\Pi_{1}&=\frac{1}{4}(-\I\ket{0}+\ket{1})(\I\bra{0}+\bra{1})\;,\\
\Pi_{2}&=\frac{1}{4}(\I\ket{0}+\ket{1})(-\I\bra{0}+\bra{1})\;,\\
\Pi_{3}&=\frac{1}{4}(\ket{0}+\ket{1})(\bra{0}+\bra{1})\;,\\
\Pi_{4}&=\frac{1}{4}(-\ket{0}+\ket{1})(-\bra{0}+\bra{1})\;.
\end{split}
\end{equation}
The probability of each of the four outcomes is $\frac{1}{4}$. By attaching an estimator coefficient $\xi_{j,k}$ to each measurement outcome, it is possible to construct unbiased estimators for the parameters we want to sense, $\hat{\theta}_{j}=\sum_{k}p_k\xi_{j,k}$, where $p_k$ is the probability of the $k$th POVM outcome occuring. From this particular POVM, we can construct unbiased estimators using the following estimator coefficients $\xi_{x,2}=\xi_{y,3}=-\xi_{x,1}=-\xi_{y,4}=2/(1-\epsilon)$ and $\xi_{i,j}=0$ for all other $i$ and $j$. These estimators then give individual variances of 
\begin{align}
v_x=\frac{1}{4}(\xi_{x,1}^2+\xi_{x,2}^2)\;,\\
v_y=\frac{1}{4}(\xi_{y,3}^2+\xi_{y,4}^2)\;,
\end{align}
which gives a total variance of $v_x+v_y=4/(1-\epsilon)^2$ coinciding with the single-copy Nagaoka bound $\mathcal{N}_{1}$, Eq.~\eqref{Nag12}. Alternatively, from the POVM, it is possible to compute the classical Fisher information without using the estimator coefficients.

This measurement strategy is theoretically optimal when measuring probe states individually. However, it requires four measurement outcomes, meaning it is necessary to use an ancilla qubit to implement this POVM. In this case Naimark's theorem can be used to convert the POVM to a projective measurement in a higher dimensional Hilbert space~\cite{neumark1943spectral}, but this comes at the cost of increasing experimental complexity. This can be avoided by noting that this POVM is equivalent to measuring $\sigma_x$ half of the time and $\sigma_y$ half of the time, where $\sigma_x$ and $\sigma_y$ are the usual Pauli matrices. Thus we can simply split our measurement in two. We measure $\theta_x$ with half of the probe states using the following POVM
\begin{equation}
\begin{split}
\label{POVM1qDCthx}
\Pi_{x1}&=\frac{1}{2}(-\I\ket{0}+\ket{1})(\I\bra{0}+\bra{1})\;,\\
\Pi_{x2}&=\frac{1}{2}(\I\ket{0}+\ket{1})(-\I\bra{0}+\bra{1})\;,
\end{split}
\end{equation}
and $\theta_y$ with the remaining probe states using
\begin{equation}
\begin{split}
\label{POVM1qDCthy}
\Pi_{y1}&=\frac{1}{2}(\ket{0}+\ket{1})(\bra{0}+\bra{1})\;,\\
\Pi_{y2}&=\frac{1}{2}(-\ket{0}+\ket{1})(-\bra{0}+\bra{1})\;.
\end{split}
\end{equation}

The new estimator coefficients are reduced by a factor of two $\xi_{x,2}^*=\xi_{y,1}^*=-\xi_{x,1}^*=-\xi_{y,2}^*=1/(1-\epsilon)$, the probability of each outcome is now $\frac{1}{2}$ and each estimate uses half as many resources. Therefore the variances in estimating $\theta_x$ and $\theta_y$ are given by
\begin{align}
v_x=2(\frac{1}{2})((\xi_{x,1}^{*})^2+(\xi_{x,2}^{*})^2)=\frac{1}{4}(\xi_{x,1}^2+\xi_{x,2}^2)\;,\\
v_y=2(\frac{1}{2})((\xi_{y,1}^{*})^2+(\xi_{y,2}^{*})^2)=\frac{1}{4}(\xi_{y,3}^2+\xi_{y,4}^2)\;,
\end{align}
coinciding with the original measurement, Eq.~\eqref{POVM1qDC}.

For the two-copy measurement we write the POVM as $\Pi_j=\ket{\psi_j}\bra{\psi_j}$ where
\begin{equation}
\label{twocopyproj}
\ket{\psi_1}=\begin{pmatrix}
\frac{1}{\sqrt{3}}\\
\frac{-\text{exp}(\mathrm{i}\pi/4)}{\sqrt{6}}\\
\frac{-\text{exp}(\mathrm{i}\pi/4)}{\sqrt{6}}\\
\frac{\text{exp}(\mathrm{i}\pi/2)}{\sqrt{3}}
\end{pmatrix}\;, \qquad
\ket{\psi_2}=\begin{pmatrix}
\frac{1}{\sqrt{3}}\\
\frac{\text{exp}(-\mathrm{i}\pi/12)}{\sqrt{6}}\\
\frac{\text{exp}(-\mathrm{i}\pi/12)}{\sqrt{6}}\\
\frac{\text{exp}(-\mathrm{i}\pi/6)}{\sqrt{3}}
\end{pmatrix}\;, \qquad
\ket{\psi_3}=\begin{pmatrix}
\frac{1}{\sqrt{3}}\\
\frac{\text{exp}(\mathrm{i}7\pi/12)}{\sqrt{6}}\\
\frac{\text{exp}(\mathrm{i}7\pi/12)}{\sqrt{6}}\\
\frac{\text{exp}(-\mathrm{i}5\pi/6)}{\sqrt{3}}
\end{pmatrix}\;,\qquad
 \ket{\psi_4}=\begin{pmatrix}
0\\
\frac{1}{\sqrt{2}}\\
\frac{-1}{\sqrt{2}}\\
0
\end{pmatrix}\;. 
\end{equation}
The first three POVM outcomes occur with probability $(4+\epsilon(\epsilon-2))/12$ and the fourth outcome occurs with probability $\epsilon(2-\epsilon)/4$. We use the following estimator coefficients
\begin{align}
\xi_{x,1}=\frac{1}{1-\epsilon},\qquad\xi_{x,2}=\frac{\sqrt{3}-1}{2(1-\epsilon)},\qquad \xi_{x,3}=\frac{-\sqrt{3}-1}{2(1-\epsilon)}\qquad\text{and}\qquad\xi_{x,4}=0\\
\xi_{y,1}=\frac{-1}{1-\epsilon},\qquad\xi_{y,2}=\frac{\sqrt{3}+1}{2(1-\epsilon)},\qquad \xi_{y,3}=\frac{1-\sqrt{3}}{2(1-\epsilon)}\qquad\text{and}\qquad\xi_{y,4}=0\;.
\end{align}
The individual variances are then given by
\begin{align}
v_x=\frac{1}{4}+\frac{3}{4(1-\epsilon)^2}\;,\\
v_y=\frac{1}{4}+\frac{3}{4(1-\epsilon)^2}\;,
\end{align}
and the sum coincides with the two-copy Nagaoka bound, Eq.~\eqref{Nag22}. A geometrical interpretation of this POVM is provided in~\ref{SN:probsimp}.

\subsection{Optimised three- and four-copy projective measurements surpassing the preceding Nagaoka bound}
\label{SN:threecopyPOVM}
In this Supplementary Note we present the details of the three- and four-copy measurements mentioned in the main text, which surpass the two- and three-copy Nagaoka bounds respectively. A three-copy POVM and estimator function which surpass the two-copy Nagaoka bound were found numerically for $\epsilon=0.5$. For this particular decoherence strength, the Holevo bound on the variance when measuring $\theta_x$ and $\theta_y$, per qubit, is 12 rad$^2$. The single-, two- and three-copy Nagaoka bounds on the variance are 16 rad$^2$, 13 rad$^2$ and 12.716 rad$^2$ respectively. Our specific three-copy measurement is theoretically able to attain a variance of 12.719, almost the same as the three-copy Nagaoka bound. We use the following 8 POVM elements to obtain this variance.
\begin{equation}
\begin{split}
\hspace{-1.75em}
&\ket{\psi_1^\text{3c}}=\begin{pmatrix}
0.4993 + 0.0260\mathrm{i}\\
-0.2887 + 0.0006\mathrm{i}\\
-0.2887 + 0.0006\mathrm{i}\\
0.2882 - 0.0162\mathrm{i}\\
-0.2887 + 0.0006\mathrm{i}\\
0.2882 - 0.0162\mathrm{i}\\
0.2882 - 0.0162\mathrm{i}\\
-0.4970 + 0.0551\mathrm{i}
\end{pmatrix},\hspace{0.25em}
\ket{\psi_2^\text{3c}}=\begin{pmatrix}
0.4968 - 0.0563\mathrm{i}\\
0.2847 - 0.0480\mathrm{i}\\
0.2847 - 0.0480\mathrm{i}\\
0.2816 - 0.0633\mathrm{i}\\
0.2847 - 0.0480\mathrm{i}\\
0.2816 - 0.0633\mathrm{i}\\
0.2816 - 0.0633\mathrm{i}\\
0.4812 - 0.1359\mathrm{i}
\end{pmatrix},\hspace{0.25em}
\ket{\psi_3^\text{3c}}=\begin{pmatrix}
0.4086 - 0.2881\mathrm{i}\\
0.1789 + 0.2266\mathrm{i}\\
0.1789 + 0.2266\mathrm{i}\\
-0.2166 + 0.1909\mathrm{i}\\
0.1789 + 0.2266\mathrm{i}\\
-0.2166 + 0.1909\mathrm{i}\\
-0.2166 + 0.1909\mathrm{i}\\
-0.3504 - 0.3566\mathrm{i}
\end{pmatrix},\\ \hspace{-1.25em}
&\ket{\psi_4^\text{3c}}=\begin{pmatrix}
0\\
-0.0076 - 0.0237\mathrm{i}\\
-0.4459 - 0.2057\mathrm{i}\\
-0.4754 + 0.1796\mathrm{i}\\
0.4535 + 0.2294\mathrm{i}\\
0.4528 - 0.1900\mathrm{i}\\
0.0226 + 0.0104\mathrm{i}\\
0
\end{pmatrix}, \hspace{0.25em}
\ket{\psi_5^\text{3c}}=\begin{pmatrix}
0\\
0.0015 - 0.0248\mathrm{i}\\
-0.3411 - 0.3532\mathrm{i}\\
0.5082 + 0.0049\mathrm{i}\\
0.3396 + 0.3781\mathrm{i}\\
-0.4909 + 0.0130\mathrm{i}\\
-0.0173 - 0.0179\mathrm{i}\\
0
\end{pmatrix},\hspace{0.25em}
\ket{\psi_6^\text{3c}}=\begin{pmatrix}
0\\
-0.4038 + 0.4119\mathrm{i}\\
0.2011 - 0.2275\mathrm{i}\\
0.2740 + 0.0008\mathrm{i}\\
0.2026 - 0.1844\mathrm{i}\\
0.3020 - 0.0320\mathrm{i}\\
-0.5760 + 0.0313\mathrm{i}\\
0
\end{pmatrix},\\ \hspace{-1.25em}
&\ket{\psi_7^\text{3c}}=\begin{pmatrix}
0\\
0.5637 + 0.1225\mathrm{i}\\
-0.2997 - 0.0491\mathrm{i}\\
0.1452 + 0.2323\mathrm{i}\\
-0.2640 - 0.0734\mathrm{i}\\
0.1879 + 0.2386\mathrm{i}\\
-0.3331 - 0.4709\mathrm{i}\\
0
\end{pmatrix},\hspace{0.25em}
\ket{\psi_8^\text{3c}}=\begin{pmatrix}
0.0574 + 0.4967\mathrm{i}\\
 0.2846 - 0.0486\mathrm{i}\\
0.2846 - 0.0486\mathrm{i}\\
-0.0640 - 0.2815\mathrm{i}\\
0.2846 - 0.0486\mathrm{i}\\
-0.0640 - 0.2815\mathrm{i}\\
-0.0640 - 0.2815\mathrm{i}\\
-0.4809 + 0.1370\mathrm{i}
\end{pmatrix}\;.
\end{split}
\end{equation}
 
These POVMs, combined with the following estimator coefficients
\begin{equation}
\begin{split}
&\xi_{x1}=-0.1338, \hspace{0.5em} \xi_{x2}=0.1338, \hspace{0.5em} \xi_{x3}=-2.4681, \hspace{0.5em} \xi_{x4}=0.7817,\\ &\xi_{x5}=-0.7817,  \hspace{0.5em}\xi_{x6}=0.7156, \hspace{0.5em} \xi_{x7}=-0.7156, \hspace{0.5em} \xi_{x8}=2.4681
\end{split}
\end{equation}
and
\begin{equation}
\begin{split}
&\xi_{y1}=-2.4680, \hspace{0.5em} \xi_{y2}=2.4680, \hspace{0.5em} \xi_{y3}=0.1338, \hspace{0.5em} \xi_{y4}=0.7157,\\& \xi_{y5}=-0.7157, \hspace{0.5em} \xi_{y6}=-0.7819, \hspace{0.5em} \xi_{y7}=0.7819, \hspace{0.5em} \xi_{y8}=-0.1338\;,
\end{split}
\end{equation}
give rise to a variance of 12.719. There is an extremely small difference of 0.003 rad$^2$ in the variance of this measurement and the three-copy Nagaoka bound, but this is not due to numerical error. Indeed, we were able to find a 10 outcome POVM with a variance equal to the three-copy Nagaoka bound (up to numerical error $\approx10^{-6}$). However, in the main text, the near-optimal 8 outcome POVM was implemented, because this does not require any ancilla qubits, simplifying the experimental realisation.

In the main text we also presented simulations based on a projective four-copy measurement, which theoretically surpasses the three-copy Nagaoka bound. The details of this POVM are found in our publicly available repository~\cite{Conlon2022}. The four-copy Nagaoka bound on the variance is  12.368 rad$^2$. Our four-copy projective measurement attains a variance of 12.508 rad$^2$, which is below the three-copy Nagaoka bound, but does not saturate the four-copy Nagaoka bound. Due to computational restraints, we only searched for projective four-copy measurements, so saturating the four-copy Nagaoka bound may still be possible with a more general four-copy POVM.

\subsection{Experimental implementation of optimal POVMs}
\subsubsection{Single- and two-copy POVM implementation}
In this Supplementary Note we describe how to convert the optimal POVMs into experimentally realisable quantum circuits. The quantum processors used in our experiments make measurements in the $z$ basis. Therefore, in order to implement the POVMs it is necessary to diagonalise them in this basis. For example, for the two qubit POVM, we aim to find a unitary matrix $U$ such that
\begin{equation}
U\begin{bmatrix}
\ket{\psi_1}&\ket{\psi_2}&\ket{\psi_3}&\ket{\psi_4}\\
\end{bmatrix}\;,
\end{equation}
is diagonal in the computational basis, where $\begin{bmatrix}
\ket{\psi_1}&\ket{\psi_2}&\ket{\psi_3}&\ket{\psi_4}\\
\end{bmatrix}$ is a $4\times4$ matrix. Implementing this unitary matrix and measuring in the $z$ basis is equivalent to implementing the desired POVM.

We start with the single qubit POVM for estimating $\theta_x$, Eq.~\eqref{POVM1qDCthx}. We write 
 \begin{equation}
\ket{\psi_{x1}}=\frac{1}{\sqrt{2}}(-\I\ket{0}+\ket{1})\qquad\text{and}\qquad \ket{\psi_{x2}}=\frac{1}{\sqrt{2}}(\I\ket{0}+\ket{1})\;,
\end{equation}
so that we wish to diagonalise $\begin{bmatrix}\ket{\psi_{x1}}&\ket{\psi_{x2}}\end{bmatrix}$. It is easy to verify that this is diagonalised by the following unitary matrix
\begin{equation}
U_x=\frac{1}{\sqrt{2}}\begin{pmatrix}
\mathrm{i}&1\\
-\mathrm{i}&1
\end{pmatrix}\;.
\end{equation}
 
In Fig.~1 of the main text, single qubit unitary matrices are shown as green boxes in the circuit diagram. We use the following definition for the most general single qubit gate characterised by three parameters, $\theta,\phi$ and $\lambda$
\begin{equation}
\label{singlequnitary}
U_{\text{1q}}(\theta,\phi,\lambda)=\begin{pmatrix}
\text{cos}(\frac{\theta}{2})&-e^{\mathrm{i}\lambda}\text{sin}(\frac{\theta}{2})\\
e^{\mathrm{i}\phi}\text{sin}(\frac{\theta}{2}) & e^{\mathrm{i}(\lambda+\phi)}\text{cos}(\frac{\theta}{2})
\end{pmatrix}\;.
\end{equation}
With this definition, up to a global phase factor, we can write $U_x=U_{\text{1q}}(3\pi/2,0,3\pi/2)$. Following the same approach, to implement the POVM for estimating $\theta_y$, Eq.~\eqref{POVM1qDCthy}, we require the following unitary matrix
\begin{equation}
U_y=\frac{1}{\sqrt{2}}\begin{pmatrix}
1&1\\
-1&1
\end{pmatrix}\;.
\end{equation}
This can be implemented as $U_y=U_{\text{1q}}(3\pi/2,0,0)=R_y(3\pi/2)$.

Optically these unitary matrices are implemented through a motorised quarter-wave plate (QWP) followed by a half-wave plate (HWP) and then another QWP. The HWP and QWP set to a specific angle $\theta$ implement the following unitary transformations
\begin{equation}
\text{HWP}(\theta)=e^{-\mathrm{i}\pi/2}\begin{pmatrix}
\text{cos}(\theta)^2-\text{sin}(\theta)^2&2\text{cos}(\theta)\text{sin}(\theta)\\
2\text{cos}(\theta)\text{sin}(\theta)&-\text{cos}(\theta)^2-\text{sin}(\theta)^2
\end{pmatrix}\;,
\end{equation}
and
\begin{equation}
\text{QWP}(\theta)=e^{-\mathrm{i}\pi/4}\begin{pmatrix}
\text{cos}(\theta)^2+\mathrm{i}\text{sin}(\theta)^2&(1-\mathrm{i})\text{cos}(\theta)\text{sin}(\theta)\\
(1-\mathrm{i})\text{cos}(\theta)\text{sin}(\theta)&\mathrm{i}\text{cos}(\theta)^2+\text{sin}(\theta)^2
\end{pmatrix}\;.
\end{equation}
To implement $U_x$, we use the following angles QWP($-\pi/4$)HWP($-3\pi/4$)QWP($\pi/2$), which is equivalent to $U_x$ up to a global phase shift. To implement $U_y$ we use QWP(0.788913939)HWP(-1.174581359)QWP(0.003515755). This is not an exact implementation, but the numerical error in the implementation of $U_y$ is significantly smaller than the expected MSE, hence can be ignored.

For the two-copy POVM, Eq.~\eqref{twocopyproj}, note that $\bra{\psi_1}$, $\bra{\psi_2}$, $\bra{\psi_3}$ and $\bra{\psi_4}$ are orthonormal hence the unitary matrix which diagonalises $[\ket{\psi_1},\ket{\psi_2},\ket{\psi_3},\ket{\psi_4}]$ is $[\bra{\psi_1},\bra{\psi_2},\bra{\psi_3},\bra{\psi_4}]^T$. Any two qubit unitary matrix can be implemented by three CNOT gates, four arbitrary single qubit unitary matrices, Eq.~\eqref{singlequnitary}, and three single qubit rotations~\cite{vatan2004optimal}. This is the circuit shown in Fig.~1 (f) in the main text. The rotation parameters needed to implement this unitary matrix to an accuracy of 1 part in 10 million are given in Tables.~\ref{unittab2} and \ref{unittab2rot}.
\begin{table}[h]
\begin{tabular}{|l|c|c|c|c|}\hline
&$U_1$ & $U_2$ &$U_3$ &$U_4$  \\ \hline
$\theta$&1.332047081999419 &1.332047066929274&2.328216659679420&0.813375851824116\\ \hline
$\phi$& 0.896361779524331 &0.674434373615608&1.509056929793133&0.461859523404381\\ \hline
$\lambda$& 0.228222830422120&1.342573324008889&2.257150772710729& -0.686354414697494 \\ \hline
\end{tabular}
\caption{\label{unittab2} Parameters required to implement the arbitrary single qubit unitary matrices for the optimal two-copy circuit in Fig.~1 (f) from the main text.}
\end{table}
\begin{table}[h]
\begin{tabular}{|l|c|c|c|}\hline
& $R_y$ & $R_y$ & $R_z$  \\ \hline
$\theta$&0.505232912711076&-0.705293075042498&1.378157349320011\\ \hline
\end{tabular}
\caption{\label{unittab2rot} Parameters required to implement the single qubit rotations for the optimal two-copy circuit in Fig.~1 (f) from the main text.}
\end{table}

\subsubsection{State preparation}
The final step before running these optimal measurement circuits is to prepare the correct probe state. For a given decoherence strength $\epsilon$ and rotation angles $\theta_x$ and $\theta_y$ we need to prepare three different states per qubit with the appropriate probabilities. The state $\ket{\psi_\theta}=R_y(\theta_y)R_x(\theta_x)\ket{0}$ is prepared with probability $1-\epsilon$ and the states $\ket{0}$ and $\ket{1}$ are prepared with probability $\epsilon/2$ for each qubit. For implementing collective measurements, each two qubit state is a Kronecker product of two states and is prepared with a probability equal to the product of the probabilities for each individual state. For example, the state $\ket{\psi_\theta}\otimes\ket{0}$ is prepared with probability $(1-\epsilon)\epsilon/2$. 

In principle, there should be no error in the value of $\epsilon$ used in our experiments, as the exact probability with which we prepare each state is known. However, in practice, due to non-zero state preparation error, there will be some error in the value of $\epsilon$ used. We now estimate the error in $\epsilon$, $\sigma_\epsilon$, given a certain state preparation error. We ignore readout error when calculating $\sigma_\epsilon$ as this does not affect the actual state we prepare. For simplicity, we will consider the case when $\theta_x=\theta_y=0$, so that the state we wish to prepare is $(1-\epsilon/2)\ket{0}\bra{0}+\epsilon/2\ket{1}\bra{1}$. Using $N$ qubits for the complete experiment, the true $\epsilon$ value is given by
 \begin{equation}\epsilon=2-\frac{2N_{0,\text{prep}}}{N}\;,  \end{equation}
where $N_{0,\text{prep}}$ is the number of $\ket{0}$ states which are actually prepared. We assume that the state preparation error is symmetric for preparation of both the $\ket{0}$ and $\ket{1}$ state, which we denote $p_p$. This is the probability of initialising the wrong qubit state. We can then write $\epsilon$ in terms of the number of $\ket{0}$ and $\ket{1}$ states which would ideally be prepared and the probability of incorrect initialisation
\begin{equation}\epsilon=2-\frac{2N(1-\epsilon_\text{i}/2)(1-p_p)}{N}-\frac{2Np_p\epsilon_\text{i}/2}{N}\;,  \end{equation}
where $\epsilon_\text{i}$ is the $\epsilon$ value which would be prepared in an ideal experiment. Finally, assuming that the $\ket{0}$ and $\ket{1}$ initialisation errors are independent, we find the variance of $\epsilon$ as 
\begin{equation}\sigma_\epsilon^2=\frac{4}{N}p_p(1-p_p)\;. \end{equation}
As we average our results over 400 repetitions of the same experiment, $\sigma_\epsilon^2$ is decreased by a further factor of 400. Using the calibration data for our devices, we find that $\sigma_\epsilon$ is on the order $10^{-4}$ or smaller for all devices, hence horizontal error bars on Fig 2 (e) in the main text would be negligible and we do not include them.

\begin{figure*}[t]
\includegraphics[width=\textwidth]{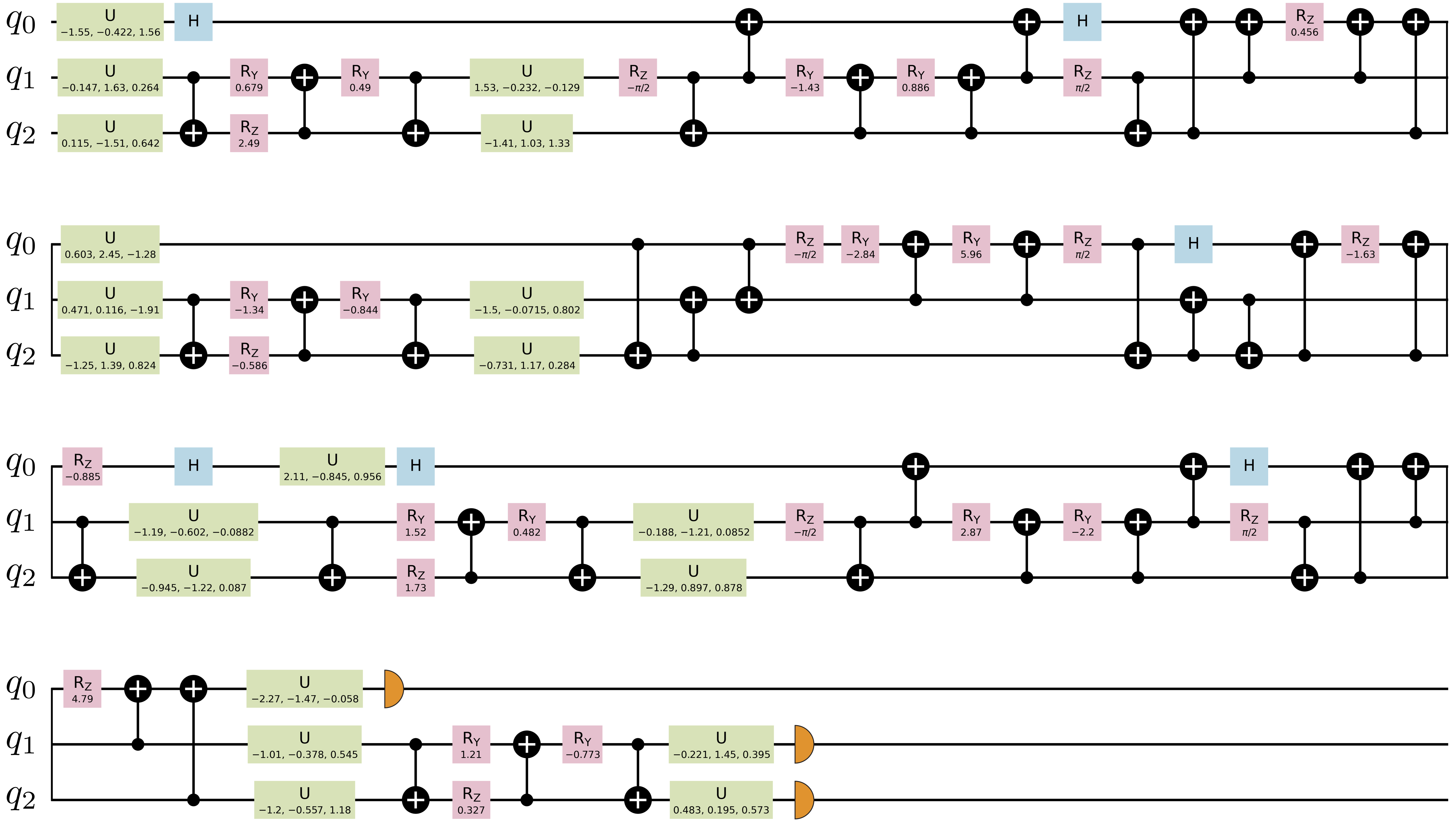}
\caption{\textbf{Three-copy measurement circuit which surpasses the two-copy Nagaoka bound.} Circuit reads from left to right and continues from top to bottom. In total the circuit requires 43 CNOT gates, 20 arbitrary single qubit unitary matrices, 6 Hadamard gates and 28 single qubit rotations, a total of 88 free parameters to optimise. The gates labelled U are as defined in Eq.~\eqref{singlequnitary}. The circuit diagram was constructed using the IBM Q QISKIT package.}
\label{fig:3copycirc}
\end{figure*}
\subsubsection{Three- and four-copy POVM implementation and noisy simulation}
 In order to implement the three-copy POVM presented in the \ref{SN:threecopyPOVM} we need a unitary matrix which diagonalises $\begin{bmatrix}\ket{\psi_1^\text{3c}},\ket{\psi_2^\text{3c}},\ket{\psi_3^\text{3c}},\ket{\psi_4^\text{3c}},\ket{\psi_5^\text{3c}},\ket{\psi_6^\text{3c}},\ket{\psi_7^\text{3c}},\ket{\psi_8^\text{3c}}\end{bmatrix}$. This POVM is constructed such that the $\ket{\psi_i^\text{3c}}$ are orthonormal. Therefore, the unitary matrix to be implemented experimentally is 
 $$\begin{bmatrix}\ket{\psi_1^\text{3c}},\ket{\psi_2^\text{3c}},\ket{\psi_3^\text{3c}},\ket{\psi_4^\text{3c}},\ket{\psi_5^\text{3c}},\ket{\psi_6^\text{3c}},\ket{\psi_7^\text{3c}},\ket{\psi_8^\text{3c}}\end{bmatrix}^\dagger.$$  Using the technique presented in Ref.~\cite{vatan2004realization}, we are able to convert this unitary matrix to a three qubit quantum circuit, shown in Fig.~\ref{fig:3copycirc}. This circuit consists of 43 CNOT gates, 20 single qubit unitary matrices, 6 Hadamard gates and 28 single qubit rotations, and is an extensive circuit. When implemented on various superconducting processors, the results of this measurement did not reach the theoretical limits. This can possibly be attributed to the gate error rates of the different processors or crosstalk between qubits. The quantum circuit corresponding to our four-copy POVM can be found in our publicly available repository~\cite{Conlon2022}. The decomposition of the four-copy measurement which we use contains 115 CNOT gates, hence unsurprisingly the four-copy POVM cannot reach the theoretical limits with even small amounts of noise. In Fig.~2 (f) in the main text we simulated implementing these measurements on noisy quantum computers. This simulation used the IBM Q QISKIT package noise models and noise was modelled as depolarising noise for different gate error rates. The results of this simulation suggest that with realistic future noise levels, quantum processors may be able to implement three-copy measurements with a precision approaching the theoretical limits. Four-copy measurements, on the other hand, will require considerably lower gate error rates to reach the theoretical limits. Although this discussion comes with the caveat that modelling noise in such a complex system is unlikely to be overly accurate, we do observe good qualitative agreement between simulation and experiment for the three-copy measurement.

\begin{figure*}[t]
\includegraphics[width=\textwidth]{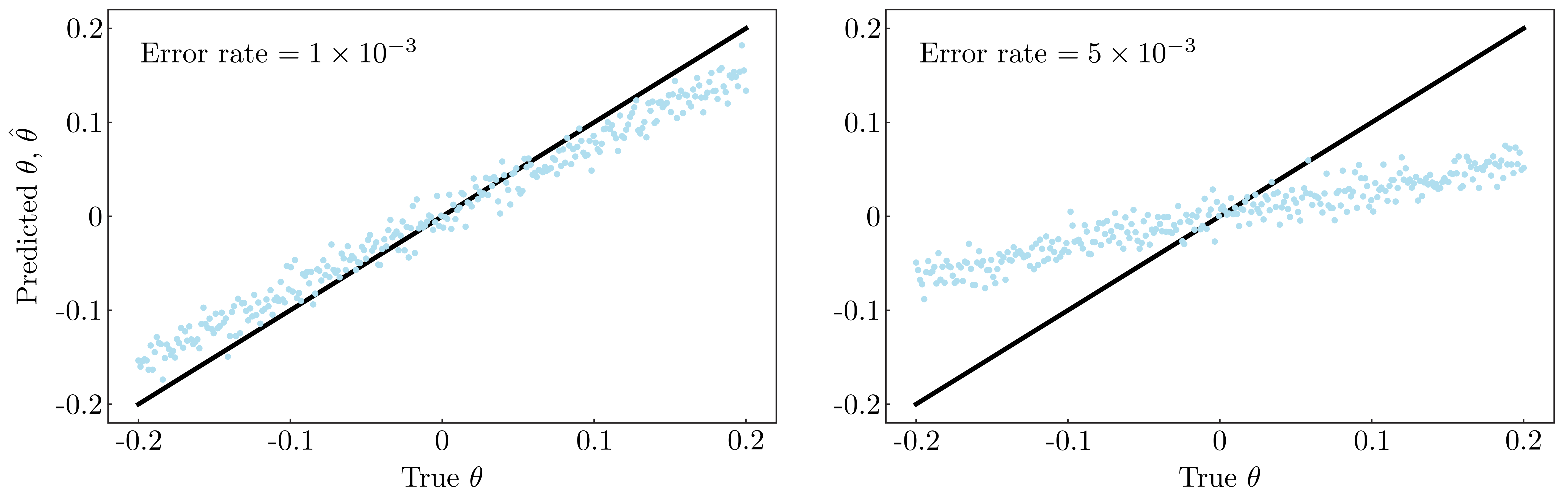}
\caption{\textbf{Noisy simulation of our four-copy measurement under depolarising noise.} Shown are the predicted values of $\theta_x$ using our four-copy measurement as a function of the true $\theta$, simulated for two different gate error rates. The black line shows the true $\theta$ value.}
\label{fig:4copygrad} 
\end{figure*}

The results presented for the noisy simulations in Fig.~2 (f) in the main text use a slightly different error mitigation method than that used for the experimental data. This is done to avoid erroneously predicting small variances from our noisy simulations. In Fig.~\ref{fig:4copygrad} we plot the predicted values of $\theta_x$ using our four-copy measurement as a function of the input $\theta$, simulated for two different gate error rates. Qualitatively, it is clear that as the noise is increasing the estimator is predicting values closer to 0. Hence, we would expect such an estimator to perform worse with increasing gate error rate. Quantitatively however, without any error mitigation, the MSE actually decreases with increasing gate error rate for $\theta=0$. With gate error rates of $1\times10^{-3}$, the MSE is 11.89 rad$^2$, for gate error rates of $5\times10^{-3}$, the MSE is 11.71 rad$^2$. The fact that higher gate error rates give a lower MSE indicates something is not correct, but even more alarming is that both of these MSEs are below what is allowed by the Holevo bound. The reason for this is that the high gate error rate biases the estimator. For an input angle of $\theta=0$, an estimator which predicts $\hat{\theta}=0$ every time will have a MSE of 0 rad$^2$. Hence, for our noisy simulations, the calibration model we use corrects for both the gradient of the estimator and the offset, $\hat{\theta}_x=m_x\hat{\theta}_{\text{noisy},x}+c_x$. As we will stress in \ref{section:bias}, a calibration model of this form can actually bias the estimator itself. Hence, we only use this form of calibration for our noisy simulation results. This is more a reflection of the limitations of our simulations than anything else. It is likely that a more comprehensive noise model is needed to completely capture the noise of a real quantum processor. 

The results of our three- and four-copy collective measurements show the trade-off between what is gained by implementing a theoretically better measurement versus what is lost by the increased experimental complexity of such a measurement. However, this work may be viewed as one small step towards implementing collective measurements on a large number of copies of the probe state simultaneously. For comparison, we will now briefly discuss alternative approaches to implementing collective measurements. All previous approaches have been restricted to implementing collective measurements on two copies of the probe state~\cite{roccia2017entangling,parniak2018beating,hou2018deterministic,wu2019experimentally,yuan2020direct} and have relied on optical systems. Owing to the way that two copies of the quantum state were created in these approaches, it is difficult to extend to measurements on more than two copies of the quantum state simultaneously. Experiments using three degrees of freedom of a single photon may offer a way around this problem~\cite{wang201818}. Another possible way of implementing collective measurements on more than two copies of a quantum state simultaneously is through quantum walks. Quantum walks were originally proposed and demonstrated for implementing POVMs on single qubit states~\cite{kurzynski2013quantum,bian2015realization,zhao2015experimental} and it was a very similar technique which has enabled some of the recent demonstrations of two-copy collective measurements~\cite{hou2018deterministic,wu2019experimentally,yuan2020direct}. The theory of quantum walks as a measurement tool has been extended to POVMs on qudit states~\cite{li2019implementation}. This, combined with recent advances in optical quantum state engineering~\cite{wang2016experimental,zhong201812}, may some day allow optical approaches to collectively measure more than two copies of the quantum state simultaneously. It is likely that the continued development of collective measurements on multiple platforms will be useful.

\subsection{Effect of error mitigation on the bias of an estimator}
\label{section:bias}

\begin{figure*}[t]
\includegraphics[width=0.9\textwidth]{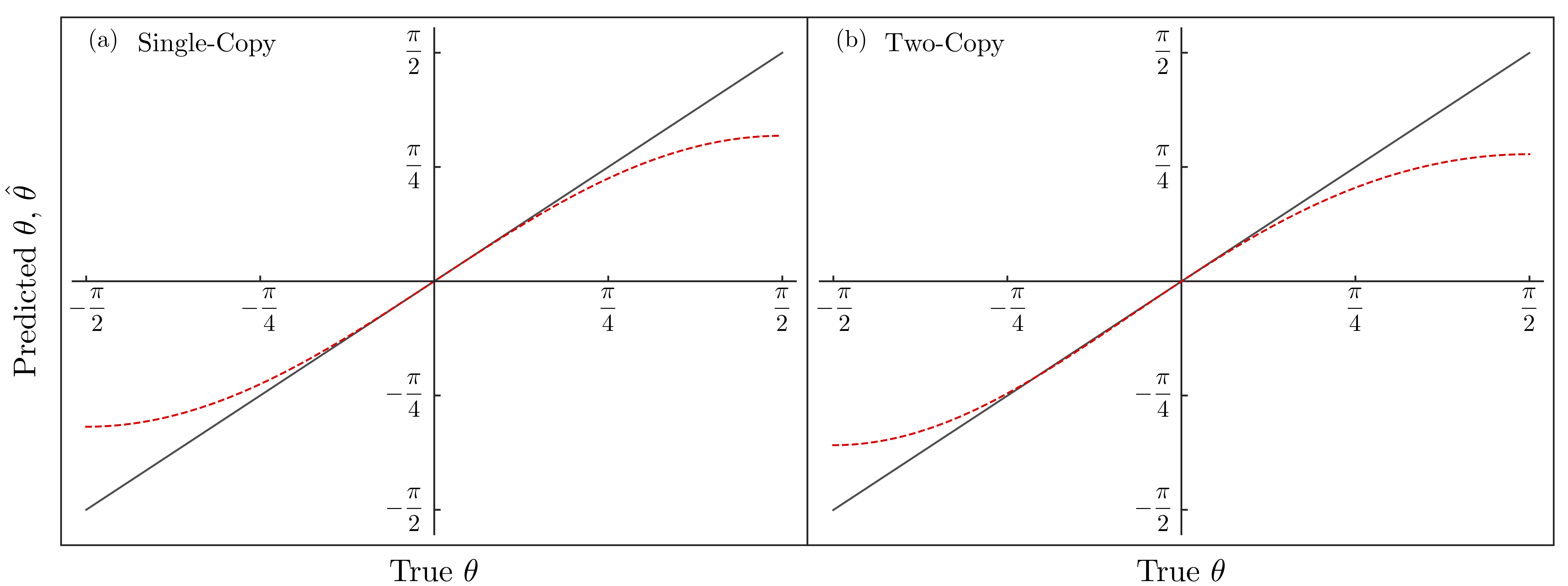}
\caption{\textbf{Predicted $\theta_x$ values.} For $\theta$ close to 0 the estimator is unbiased, i.e. $\hat{\theta}=\theta$. However, for larger $\theta$ the estimator is no longer unbiased. This is to be expected as the estimator was constructed to be unbiased in the region around $\theta=0$. (a) and (b) show results for the single-copy and two-copy estimator respectively. Results for estimating $\theta_x$ and $\theta_y$ are similar. The black line shows the true $\theta$ values.}
\label{fig:unbiased_est}
\end{figure*}

When calculating the various bounds in \ref{SN1} it was assumed that the rotations to be estimated are small, meaning the estimators are unbiased exactly at $\theta=0$. It is not guaranteed that the estimators will remain unbiased away from $\theta=0$. Fig.~\ref{fig:unbiased_est} shows the predicted angles, $\hat{\theta}$, for a range of different input angles, $\theta$. Evidently, the estimator remains approximately unbiased for a large range of $\theta$. It is important for any estimator to be unbiased to ensure a fair comparison is being made. Provided we have sufficiently many probe states available, it is always possible to operate in the region where the estimator is unbiased. This can be done by taking a small sample of the probe states, $\sqrt{N}\ll N$, where $N$ is the number of available probe states, to obtain a rough estimate of $\theta$. The measurement apparatus can then be adjusted to operate in the unbiased region by taking into account this rough estimate. 

\begin{figure*}[t]
\includegraphics[width=0.5\textwidth]{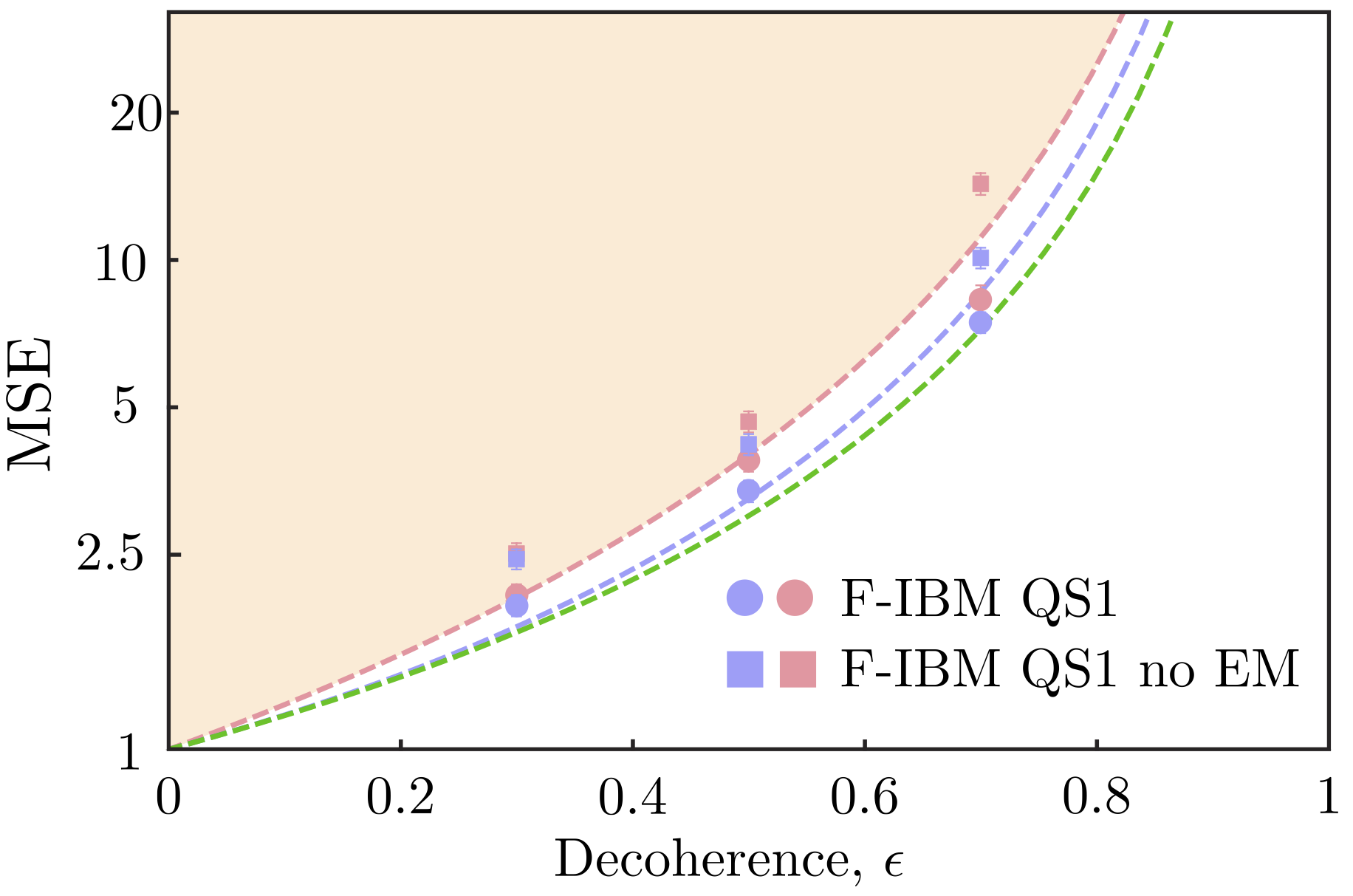}
\caption{\textbf{Effect of naive error mitigation on estimation performance.} We show the same results as presented in Fig.~2 (e) of the main text for the IBM Q System One device, however now the error mitigation (EM) is performed using the model given by Eq.~\eqref{eq:linmit}, instead of the model used in the main text, Eq.~\eqref{eq:modelmain}. Data is presented for $\epsilon=0.3,$ 0.5 and 0.7. We show the MSE obtained both with and without error mitigation. For $\epsilon=0.3$ the MSE is now considerably larger than in the main text, while for $\epsilon=0.7$ it is considerably reduced. This can be explained by the fitted value of $m_{x(y)}$. As in the main text the dashed pink, purple and green lines correspond to the single-copy Nagaoka, two-copy Nagaoka and Holevo bounds respectively. }
\label{fig:biased_est}
\end{figure*}

In this work error mitigation was used to allow us to observe quantum-enhanced metrology. An important requirement on any error mitigation technique used is that it does not introduce any bias into the estimator. We primarily focused on Clifford data regression error mitigation~\cite{czarnik2021error} which essentially amounts to producing a model for the system being interrogated. In the main text we used a model of the form
\begin{equation}
\label{eq:modelmain}
\hat{\theta}_{x(y)}=\hat{\theta}_{\text{noisy, }x(y)}+c_{x(y)}\;,
\end{equation}
where $c_{x(y)}$ is a constant which accounts for the offset of the true estimate from the noisy estimate. However, the authors who introduced this form of error mitigation originally proposed a linear model~\cite{czarnik2021error} of the form
\begin{equation}
\label{eq:linmit}
\hat{\theta}_{x(y)}=m_{x(y)}\hat{\theta}_{\text{noisy, }x(y)}+c_{x(y)}\;,
\end{equation}
where the extra term $m_{x(y)}$ is another constant, determining the slope of the model. Given that the linear model proposed is motivated by depolarising noise, it seems a natural choice. While this may be true for many other applications of quantum processors, it is not true for quantum metrology. For quantum metrology $\hat{\theta}_{\text{noisy, }x(y)}$ will be some distribution of estimated angles with a certain variance. If we were to multiply this distribution by some constant $m_{x(y)}$ which is less than 1, we would artificially reduce the variance of the estimator. It then becomes possible to have variances which appear smaller than the minimum allowed by quantum mechanics. The effect of naively applying error mitigation based on Eq.~\eqref{eq:linmit} to quantum metrology is shown in Fig.~\ref{fig:biased_est}. In this example the fitted gradient depends on the value of $\epsilon$: for larger $\epsilon$, $m$ is smaller. 
 As Fig.~\ref{fig:biased_est} shows, with this error mitigation model, we apparently surpass the two-copy Nagaoka bound in some regions, whereas, with an unbiased estimator this is not possible. 
  Thus, it is clear that error mitigation cannot be naively applied to quantum metrology. Other error mitigation techniques were investigated, including zero noise extrapolation~\cite{temme2017error,kandala2019error,li2017efficient} and quantum readout-error mitigation~\cite{maciejewski2020mitigation,mooney2021generation}, however they were not found to be as effective as Clifford data regression. Specifically, zero noise extrapolation introduces a large overhead, which is not ideal for quantum metrology, and quantum readout-error mitigation offered only marginal improvements in the MSE. Further study is needed to fully understand what error mitigation protocols are best for quantum metrology. Indeed, recent error mitigation proposals, specifically designed for quantum metrology, may prove more effective~\cite{yamamoto2021error}. 
  
\subsection{Further three-copy measurement results}
\begin{figure*}[t]
\includegraphics[width=\textwidth]{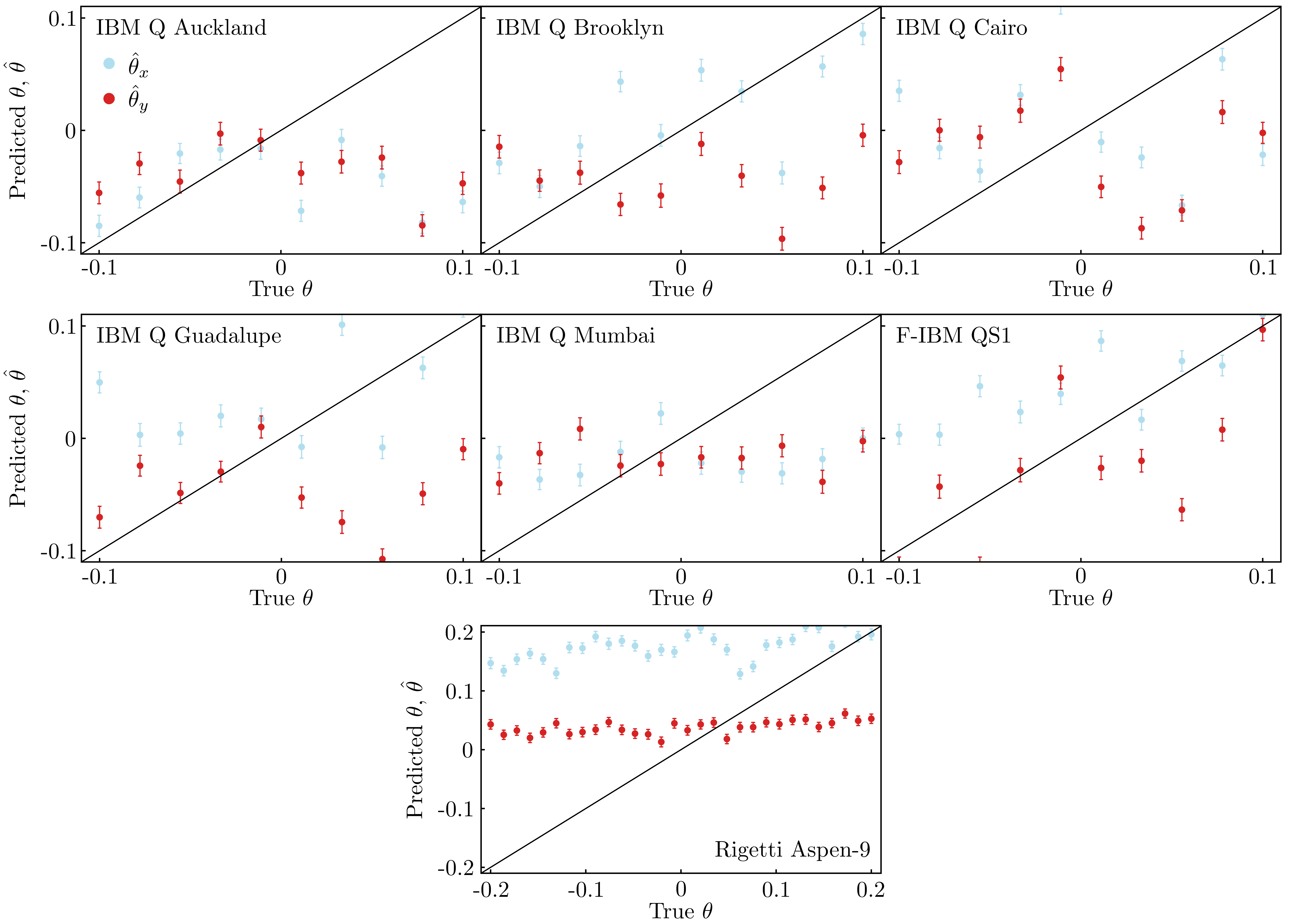}
\caption{\textbf{Three-copy measurement implemented on several quantum processors.} We show the estimated $\theta_x$ and $\theta_y$ values, predicted by the three-copy measurement, as a function of the true angle $\theta$, shown as the black line. For all devices, there is no meaningful correlation between the true and predicted values, rendering them effectively useless as estimators. All data points are based on at least 17,000 shots and all error bars correspond to one standard deviation obtained through bootstrapping.}
\label{Fig:threecopyuseless}
\end{figure*}

Fig.~2 in the main text shows the results of our three-copy measurement implemented on the Rigetti Aspen-9 and F-IBM QS1 processors. Although these measurements did not reach the theoretical limit, the MSE for the F-IBM QS1 processor is within a factor of 2 of this limit. This perhaps suggests that with minor improvements in gate error rates, the theoretical limits on three-copy measurements may be approached. However, as we now show, this is not necessarily true. We estimated a range of angles using the three-copy measurement on several different quantum processors. The results of this are shown in Fig.~\ref{Fig:threecopyuseless}, where it is evident that the three-copy measurements are effectively useless at distinguishing different angles. For all devices tested, there was no meaningful correlation between the input angle and the estimated angle. This is in stark contrast to the single-copy and two-copy measurements, shown in Figs.~1 and 2 in the main text. The data-set measured for the Rigetti Aspen-9 device differs slightly from the IBM Q devices, due to different device accessibility. Further theoretical and experimental studies will be required to fully understand the utility of three-copy collective measurements for metrology. 

\subsection{ Problem where collective measurements on many copies of the probe state are necessary}
For the problem considered in this work, two-copy measurements are able to achieve almost the same precision as the Holevo bound. Performing collective measurements on three or more copies of the probe state offers only marginal improvements in the precision at the cost of greater experimental complexity. Thus, it is natural to wonder if it is truly necessary to implement collective measurements on more than two copies of the quantum state. In this Supplementary Note we provide an example which clearly shows that collective measurements on many copies of the quantum state are necessary. 

We examine a similar problem to the main text, estimating qubit rotations subject to the amplitude damping channel, using the probe state $\ket{1}$. After the rotations and amplitude damping, the probe and its derivatives are given by 
\begin{equation}
\rho=\begin{pmatrix}
\text{p}&0\\
0&1-\text{p}
\end{pmatrix},\qquad
\partial_x\rho=\frac{\mathrm{i}\sqrt{1-\text{p}}}{2}\begin{pmatrix}
0&-1\\
1&0
\end{pmatrix},\qquad\text{and}\qquad
\partial_y\rho=\frac{\sqrt{1-\text{p}}}{2}\begin{pmatrix}
0&-1\\
-1&0
\end{pmatrix}\;,
\end{equation}
where p is the amplitude damping strength. The following matrices then optimise the Holevo and Nagaoka bounds
\begin{equation}
X_x=\frac{\mathrm{i}}{\sqrt{1-\text{p}}}\begin{pmatrix}
0&-1\\
1&0
\end{pmatrix}\qquad\text{and}\qquad
X_y=\frac{1}{\sqrt{1-\text{p}}}\begin{pmatrix}
0&-1\\
-1&0
\end{pmatrix}\;.
\end{equation}
By direct substitution it can be verified that these matrices satisfy the unbiased conditions, Eq.~\eqref{eq:unbiased}. The corresponding Holevo bound is given by
\begin{equation}
\mathcal{H}_{\text{AD}}=    \bigg\{\begin{array}{lll}
           4&\text{for}&\text{p} \leq 1/2\;,\\
           \frac{4\text{p}}{1-\text{p}}&\text{for}&\text{p} > 1/2\;.
                                      \end{array}
                                      \;.
\end{equation}
The single-copy Nagaoka bound is given by
\begin{equation}
\mathcal{N}_{1,\text{AD}}=\frac{4}{1-\text{p}}\;.
\end{equation}
 
Considering measurements on two copies of the probe state simultaneously, the following matrices optimise the two-copy Nagaoka bound
\begin{equation}
X_x=\frac{\mathrm{i}}{2\sqrt{1-\text{p}}}\begin{pmatrix}
0&-1&-1&0\\
1&0&0&-1\\
1&0&0&-1\\
0&1&1&0
\end{pmatrix}\qquad\text{and}\qquad
X_y=\frac{-1}{2\sqrt{1-\text{p}}}\begin{pmatrix}
0&1&1&0\\
1&0&0&1\\
1&0&0&1\\
0&1&1&0
\end{pmatrix}\;.
\end{equation}
The two-copy Nagaoka bound is given by
\begin{equation}
2\mathcal{N}_{2,\text{AD}}=\frac{4}{1-\text{p}}-4\text{p}\;,
\end{equation}
where we include the factor of two to account for the resources used.

\begin{figure}[t]
\includegraphics[width=\textwidth]{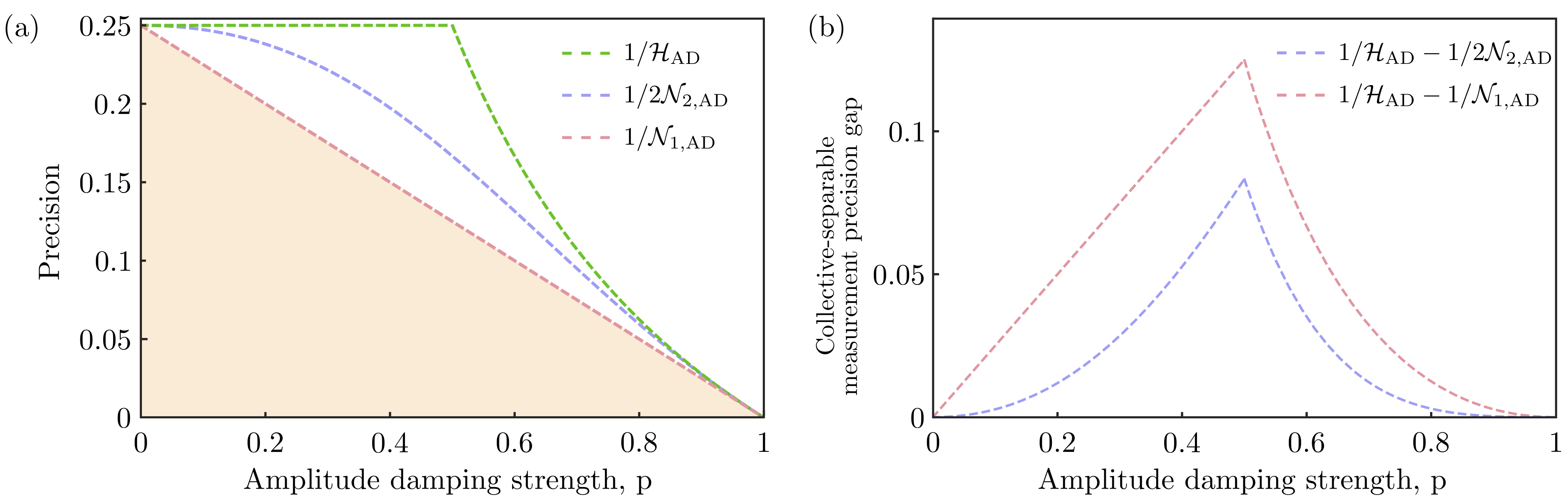}
\caption{\textbf{Attainable precision with separable and collective measurements in the amplitude damping channel.} (a) Inverse of the single-copy Nagaoka, two-copy Nagaoka and Holevo bounds as a function of the amplitude damping strength p. Although the two-copy Nagaoka bound is closer to the Holevo bound that the single-copy Nagaoka bound, there still remains a large gap between the two-copy Nagaoka and Holevo bounds. (b) The difference between the Holevo and Nagaoka bounds as a function of the amplitude damping strength. The large gap between the two-copy Nagaoka bound and the Holevo bound can only be narrowed by collective measurements on more than two copies of the probe state.}
\label{fig:adfig}
\end{figure}

The inverse of the single-copy Nagaoka, two-copy Nagaoka and the Holevo bounds are shown in Fig.~\ref{fig:adfig} (a). Although the two-copy Nagaoka bound is closer to the Holevo bound, there remains a considerable gap between these two bounds. The difference between both Nagaoka bounds and the Holevo bound is shown in Fig.~\ref{fig:adfig} (b). For this example, it is evident that measurements on more than two copies of the probe state will be required if the ultimate limits in quantum metrology are to be attained. This is also a physically relevant channel, as the amplitude damping channel can be used to model the decay of an atom from its excited state to its ground state. We can expect that many other tasks in quantum information will require collective measurements on many copies of the probe state.

\subsection{Computing Lu and Wang's metrological bound based on Heisenberg's uncertainty principle}
Recently Lu and Wang derived a trade-off relation between measurement variances when estimating two parameters based on uncertainty relations~\cite{lu2021incorporating}. As in the main text, we shall refer to this bound as the LW uncertainty relation. We now compute the LW uncertainty relation for the problem considered in the main text. To do so we first compute the symmetric logarithmic derivative (SLD) quantum Fisher information matrix $\mathcal{F}$,
\begin{equation}
\mathcal{F}_{jk}=\text{Re}(Q_{jk})=\text{Re}(\text{Tr}[L_jL_k\rho])\;,
\end{equation}
where $L_j$ satisfies $(L_j\rho+\rho L_j)/2=\partial_j\rho$. It is easily verified that the matrix $Q$ is given by
\begin{equation}
Q=\begin{pmatrix}
(1-\epsilon)^2&\mathrm{i}(1-\epsilon)^3\\
-\mathrm{i}(1-\epsilon)^3&(1-\epsilon)^2
\end{pmatrix}\;.
\end{equation}
Based on this the $\tilde{c}_{jk}$ terms (Eq. (7) of Ref.~\cite{lu2021incorporating}) can be computed as $\tilde{c}_{11}=\tilde{c}_{22}=0$ and $\tilde{c}_{12}=\tilde{c}_{21}=1$. These terms then allow the following metrological bound to be derived (Eq. (8) of Ref.~\cite{lu2021incorporating})
\begin{equation}
\frac{1}{v_x}+\frac{1}{v_y}\leq(1-\epsilon)^2\;.
\end{equation}
 
The LW uncertainty relation provides a tradeoff relation between the variances which can be attained for estimating the two rotation angles. The minimum total variance is achieved when $v_x=v_y=2/(1-\epsilon)^2$, which coincides with the single-copy Nagaoka bound. Therefore, our two-copy collective measurement, which surpasses the single-copy Nagaoka bound, can also surpass the LW uncertainty relation. To the best of our knowledge, this is the first time that a \textquote{universally valid} uncertainty principle has been surpassed, albeit indirectly, i.e. through measurement variances as opposed to directly probing the observables.

The LW uncertainty relation is based on previously derived measurement uncertainty relations which explicitly assume that only separable measurements are used. These previous bounds are based on the usual notion of uncertainty relations for operators, whereas Lu and Wang were the first to map this to quantum parameter estimation. It is possible to calculate a two-copy version of the LW uncertainty relation, however this is unsatisfying as it gives a bound which cannot be reached even with collective measurements on infinitely many copies of the probe state. In \ref{sn:adjustLW} we present one possible way to modify the LW uncertainty relation so that it accounts for collective measurements.

\subsection{POVMs violating the LW uncertainty relation with unbalanced variances}
Fig.~3 in the main text shows single-copy measurements which verify the LW relation when restricted to separable measurements for a range of $v_x$ and $v_y$ values. For $v_x=v_y$, this is achieved using the single-copy POVMs presented in \ref{SN2}, with an equal number of qubits used for estimating $\theta_x$ and $\theta_y$. For $v_x\neq v_y$ the same POVM is used, but now a different number of qubits are used for estimating $\theta_x$ and $\theta_y$. The total number of qubits used in each experiment remains fixed. By assigning more qubits to estimating $\theta_x$, we can reduce $v_x$ at the expense of increasing $v_y$ and vice versa.

For the measurements violating the LW uncertainty relation, the two-copy measurement in \ref{SN2} is sufficient for the point where $v_x=v_y$. However, for the points where $v_x\neq v_y$, a new POVM is required. The purple line in Fig.~3 is obtained from the weighted Nagaoka bound, i.e. the two-copy Nagaoka bound computed with non-identity weight matrices. Changing the weight matrix corresponds to a transformation of the parameters being estimated~\cite{fujiwara1999estimation}. Hence, after computing the weighted Nagaoka bound, we transform from the weighted variances $w_xv_x$ and $w_yv_y$ to the variances in the parameters of interest $v_x$ and $v_y$. Each different weight matrix produces a line in the $v_x-v_y$ plane. The purple curve is then the envelope of all these lines. The line corresponding to the weighted Holevo bound is calculated similarly. 

Similarly, when finding measurements where $v_x\neq v_y$, which violate the LW uncertainty relation, we find a measurement which minimises $w_xv_x+w_yv_y$. For $w_x=1.4$, $w_y=0.6$, we use the POVM $\{\Pi^{w1}_i=\ket{\psi^{w1}_{i}}\bra{\psi^{w1}_{i}}\}$, where
\begin{equation}
\begin{split}
\label{twocopyweighted1}
&\ket{\psi^{w1}_{1}}=\begin{pmatrix}
0.2813 - 0.5023\mathrm{i}\\
0.4284 + 0.0308\mathrm{i}\\
0.4284 + 0.0308\mathrm{i}\\
  0.2318 + 0.4958\mathrm{i}
  \end{pmatrix}\;, \qquad
\ket{\psi^{w1}_{2}}=\begin{pmatrix}
0\\
  -0.6664 - 0.2366\mathrm{i}\\
   0.6664 + 0.2366\mathrm{i}\\
   0
\end{pmatrix}\;, \\
&\ket{\psi^{w1}_{3}}=\begin{pmatrix}
 -0.5584 - 0.1590\mathrm{i}\\
0.3481 + 0.0991\mathrm{i}\\
0.3481 + 0.0991\mathrm{i}\\
  -0.6089 - 0.1734\mathrm{i}
\end{pmatrix}\;,\quad
 \ket{\psi^{w1}_{4}}=\begin{pmatrix}
-0.3415 + 0.4635\mathrm{i} \\
0.2048 + 0.3776\mathrm{i}\\
  0.2048 + 0.3776\mathrm{i}\\
0.5473 + 0.0066\mathrm{i}   
\end{pmatrix}\;. 
\end{split}
\end{equation}
This POVM, combined with the following estimator coefficients
\begin{equation}
\begin{split}
&\xi^{w1}_{x,1}=-2.2810 ,\qquad\xi^{w1}_{x,2}=0,\qquad \xi^{w1}_{x,3}=0\qquad\text{and}\qquad\xi^{w1}_{x,4}= 2.2810\\
&\xi^{w1}_{y,1}=1.5109,\qquad\xi^{w1}_{y,2}=0,\qquad \xi^{w1}_{y,3}= -3.1423\qquad\text{and}\qquad\xi^{w1}_{y,4}= 1.5109\;,
\end{split}
\end{equation}
gives rise to the data point in Fig.~3 with reduced variance in estimating $\theta_x$. 

Similarly, for $w_x=0.6$, $w_y=1.4$, the necessary POVM is $\{\Pi^{w2}_i=\ket{\psi^{w2}_{i}}\bra{\psi^{w2}_{i}}\}$, where
\begin{equation}
\begin{split}
\label{twocopyweighted1}
&\ket{\psi^{w2}_{1}}=\begin{pmatrix}
 0.5530 + 0.1602\mathrm{i}\\
 0.4243 - 0.0671\mathrm{i}\\
0.4243 - 0.0671\mathrm{i}\\
0.4304 - 0.3381\mathrm{i}
\end{pmatrix}\;, \qquad
\ket{\psi^{w2}_{2}}=\begin{pmatrix}
0.5515 + 0.1651\mathrm{i}\\
  -0.3202 - 0.2863\mathrm{i}\\
-0.3202 - 0.2863\mathrm{i}\\
 0.1784 + 0.5174\mathrm{i}
\end{pmatrix}\;, \\
&\ket{\psi^{w2}_{3}}=\begin{pmatrix}
0.2998 - 0.4973\mathrm{i}\\
 0.3100 + 0.1869\mathrm{i}\\
0.3100 + 0.1869\mathrm{i}\\
-0.3269 + 0.5422\mathrm{i} 
\end{pmatrix}\;,\qquad
 \ket{\psi^{w2}_{4}}=\begin{pmatrix}
0\\
-0.6643 - 0.2423\mathrm{i}\\
0.6643 + 0.2423\mathrm{i}\\
0
\end{pmatrix}\;. 
\end{split}
\end{equation}
The estimator coefficients for this POVM become
\begin{equation}
\begin{split}
&\xi^{w2}_{x,1}=1.5109 ,\qquad\xi^{w2}_{x,2}=1.5109,\qquad \xi^{w2}_{x,3}=-3.1423\qquad\text{and}\qquad\xi^{w2}_{x,4}=0\\
&\xi^{w2}_{y,1}=2.2810,\qquad\xi^{w2}_{y,2}=-2.2810,\qquad \xi^{w2}_{y,3}=0\qquad\text{and}\qquad\xi^{w2}_{y,4}=0\;.
\end{split}
\end{equation}
This measurement gives rise to the data point in Fig.~3 with reduced variance in estimating $\theta_y$.

\subsection{Measurement surpassing universally valid uncertainty relations for operators}
Surpassing Lu and Wang's bound, should be equivalent to surpassing the measurement uncertainty relations for operators on which it is based. We now show this explicitly. The LW uncertainty relation is based on an uncertainty relation which was a cumulation of work from Ozawa and Branciard~\cite{ozawa2003universally,ozawa2004uncertainty,branciard2013error,ozawa2014error}. These works provided trade-off relations for measuring two Hermitian operators $A$ and $B$. When $A$ and $B$ do not commute, they cannot be jointly measured and so instead we measure a pair of commuting observables $\mathcal{A}$ and $\mathcal{B}$ which approximate the ideal measurement. The approximate observables are measured on an extended Hilbert space of the quantum state $\rho$ combined with an ancilla state $\eta$. The measurement errors for the ideal observables $A$ and $B$ are then given by
\begin{equation}
\epsilon_A=\sqrt{\text{Tr}[(\mathcal{A}-A\otimes\mathbb{I})^2(\rho\otimes\eta)]}\quad\text{and}\quad\epsilon_B=\sqrt{\text{Tr}[(\mathcal{B}-B\otimes\mathbb{I})^2(\rho\otimes\eta)]}\;.
\end{equation}
The following uncertainty relation is then claimed to hold
\begin{equation}
\epsilon_A^2\sigma_B^2+\epsilon_B^2\sigma_A^2+2\sqrt{\sigma_A^2\sigma_B^2-D_{AB}^2}\epsilon_A\epsilon_b\geq D_{AB}^2\;,
\end{equation}
where $\sigma_{A}=\sqrt{\text{Tr}[A^2\rho]-(\text{Tr}[A\rho])^2}$ and $D_{AB}=\text{Tr}[\abs{\sqrt{\rho}(AB-BA)\sqrt{\rho}}]$/2 where $\abs{X}=\sqrt{X^\dagger X}$. In order to map our measurements to this operator approach, we need the ideal operators $A$ and $B$. For a parameter estimation problem these are always given by the SLD operators, $A=L_x$ and $B=L_y$. For the problem considered in this work, the SLD operators are
\begin{equation}
L_x=-(1-\epsilon)\sigma_y\quad\text{and}\quad L_y=(1-\epsilon)\sigma_x\;.
\end{equation}
 Substituting this in gives $\sigma_A=\sigma_B=1-\epsilon$. We can also evaluate $D_{AB}$ as $(1-\epsilon)^2$. The uncertainty relation for this problem therefore becomes
\begin{equation}
\label{eq:ucrelation}
\epsilon_A^2+\epsilon_B^2\geq (1-\epsilon)^2\;.
\end{equation}

At this point we can proceed in one of two ways to show that this relation is violated. From a metrological perspective, Lu and Wang define the \textquote{regret of the Fisher information} (hereafter abbreviated to regret), as the difference between the classical Fisher information (CFI) of a specified measurement $F$ and the SLD quantum Fisher information $\mathcal{F}$. Then an equivalence is drawn between the regret and the measurement errors for the ideal observables $R_{xx}=\epsilon_A^2$ and $R_{yy}=\epsilon_B^2$. Therefore, by computing the CFI for of our measurements we can evaluate the regret and in turn the uncertainty relation, Eq.~\eqref{eq:ucrelation}. Given a POVM $\{\Pi_m\}$ and a density matrix which depends on the parameters of interest $\rho(\theta_x,\theta_y)$, the CFI can be computed as
\begin{equation}
F_{j,k}=\sum_m\text{Tr}[\rho\Pi_m]\frac{\partial \text{log}(\text{Tr}[\Pi_m\rho])}{\partial\theta_j}\frac{\partial \text{log}(\text{Tr}[\Pi_m\rho])}{\partial\theta_k}\;,
\end{equation}
where log refers to the natural logarithm. Using the single- and two-copy measurements specified previously, Eqs.~\eqref{POVM1qDC} and \eqref{twocopyproj} respectively, we can evaluate the CFI for our two measurements as
\begin{equation}
\text{CFI}_{\rho}=\frac{(1-\epsilon)^2}{2}\begin{pmatrix}
1&0\\
0&1
\end{pmatrix}\;,
\end{equation}
and
\begin{equation}
\text{CFI}_{\rho\otimes\rho}=\frac{2(1-\epsilon)^2}{4-2\epsilon+\epsilon^2}\begin{pmatrix}
1&0\\
0&1
\end{pmatrix}\;.
\end{equation}
Note that the two-copy CFI has been scaled by a factor of two to account for the fact that twice as many resources are being used. Evaluating the regret and substituting into Eq.~\eqref{eq:ucrelation}, shows that the uncertainty relation is saturated for the single-copy measurement and violated for the two-copy measurement.

Equivalently, we can give the exact forms for the approximate observables $\mathcal{A}$ and $\mathcal{B}$. For this we require the measurement channel $\Phi:S(\mathcal{H}_S)\rightarrow S(\mathcal{H}_R)$ introduced by Lu and Wang, which maps from the set of all density matrices on a Hilbert space $\mathcal{H}_S$ to density matrices on an alternative Hilbert space $\mathcal{H}_R$ which acts as a register of all the measurement outcomes. This channel is defined as
\begin{equation}
\Phi:\rho\mapsto \sum_m\text{Tr}[\rho\Pi_m]\ket{m}\bra{m}\;,
\end{equation}
where $\ket{m}$ are states forming an orthonormal basis in $\mathcal{H}_R$. The new SLD operator for the density matrix $\Phi(\rho)$ is given by
\begin{equation}
\tilde{L}_j =\sum_m\frac{\partial\text{Tr}[\rho\Pi_m]}{\partial\theta_j}\ket{m}\bra{m}\;,
\end{equation}
 
The approximate observables can then be defined as $\mathcal{A}=U^\dagger(\mathbb{I}_S\otimes\tilde{L}_x\otimes\mathbb{I}_R)U$ and $\mathcal{B}=U^\dagger(\mathbb{I}_S\otimes\tilde{L}_y\otimes\mathbb{I}_R)U$, where $\mathbb{I}_S$ and $\mathbb{I}_R$ are the identity matrices on the system and register Hilbert spaces respectively. $U$ is a unitary matrix which satisfies
\begin{equation}
\label{eq:unitcond}
\Phi(\rho)=\text{Tr}_{1,3}[U(\rho\otimes\eta\otimes\eta)U^\dagger]\;,
\end{equation}
for all density matrices $\rho$, where $\eta$ is any state in the register Hilbert space and Tr$_{1,3}$ denotes the partial trace over the first and third systems.

We will now present unitary matrices which satisfy Eq.~\eqref{eq:unitcond} for both the single-copy and two-copy measurements, allowing the approximate observables $\mathcal{A}$ and $\mathcal{B}$ to be obtained. For the single-copy measurement we first use the Naimark extension~\cite{neumark1943spectral} to convert the measurement in Eq.~\eqref{POVM1qDC} to projectors. One possible Naimark extension is
\begin{equation}
\ket{\psi}_{1,E}=\frac{1}{\sqrt{2}}\begin{pmatrix}
1\\
0\\
\mathrm{i}\\
0
\end{pmatrix},\quad
\ket{\psi}_{2,E}=\frac{1}{\sqrt{2}}\begin{pmatrix}
1\\
0\\
-\mathrm{i}\\
0
\end{pmatrix},\quad
\ket{\psi}_{3,E}=\frac{1}{\sqrt{2}}\begin{pmatrix}
0\\
1\\
0\\
1
\end{pmatrix}\quad\text{and}\quad
\ket{\psi}_{4,E}=\frac{1}{\sqrt{2}}\begin{pmatrix}
0\\
1\\
0\\
-1
\end{pmatrix}\;.
\end{equation}
Using this projection we define the following unitary matrix
\begin{equation}
U_\Pi=\ket{0}\bra{\psi}_{1,E}+\ket{1}\bra{\psi}_{2,E}+\ket{2}\bra{\psi}_{3,E}+\ket{3}\bra{\psi}_{4,E}\;.
\end{equation}
We next define
\begin{equation}
\begin{split}
U_1&=\mathbb{I}_2\otimes C_{1,2}\otimes\mathbb{I}_4\;,\\
U_2&=U_\Pi\otimes\mathbb{I}_8\;,\\
U_3&=C_{1,4}\otimes\mathbb{I}_2\quad\text{and}\\
U_4&=\mathbb{I}_2\otimes C_{1,4}\;,
\end{split}
\end{equation}
where $\mathbb{I}_d$ is the $d-$dimensional identity matrix and 
\begin{equation}
\begin{split}
C_{1,2}=\ket{0}\bra{0}\otimes\mathbb{I}_2+\ket{1}\bra{1}\otimes \sigma_x \quad\text{and} \\ 
C_{1,4}=\ket{0}\bra{0}\otimes\mathbb{I}_8+\ket{1}\bra{1}\otimes \mathbb{I}_4\otimes \sigma_x \;. \\ 
\end{split}
\end{equation}
The necessary single-copy unitary matrix is $U_{1\text{q}}=U_\text{swap}.U_4.U_3.U_2.U_1$, where $U_{\text{swap}}$ swaps modes 2 and 3 followed by modes 1 and 2. 

The two-copy measurement presented in Eq.~\eqref{twocopyproj} is already a projective measurement, hence we do not need to invoke Naimark's theorem. The necessary unitary matrix for the two-copy measurement is very similar to the single-copy unitary matrix, however in the definition of $U_\Pi$ we need to replace $\ket{\psi}_{j,E}$ with $\ket{\psi}_j$ from Eq.~\eqref{twocopyproj}. The unitary matrix $U_\text{swap}$ now needs to swap modes 2 and 3, followed by modes 1 and 2, followed by modes 3 and 4 and finally modes 2 and 3. For the single-copy and two-copy measurements the total ancilla systems are $\ket{x_+}\bra{x_+}\otimes\ket{z_+}\bra{z_+}\otimes\ket{z_+}\bra{z_+}\otimes\ket{z_+}\bra{z_+}$ and $\ket{x_+}\bra{x_+}\otimes\ket{z_+}\bra{z_+}\otimes\ket{z_+}\bra{z_+}$ respectively, where $\ket{z_+}=(1,0)^T$ and $\ket{x_+}=(1,1)^T/\sqrt{2}$.

By inferring the approximate observables $\mathcal{A}$ and $\mathcal{B}$ we verify that the separable measurement saturates the uncertainty relation whereas the two-copy measurement violates it. As $\mathcal{A}$ and $\mathcal{B}$ commute, the measurement of one does not disturb any subsequent measure of the other. Hence, our collective measurement violating the LW uncertainty relation can be mapped to a violation of error-disturbance type uncertainty relations~\cite{ozawa2003universally}. We note here that, for the two-copy measurement, we scale the errors $\epsilon_A$ and $\epsilon_B$ by a factor of two, because they are effectively estimating the optimal operator twice. This rescaling by a factor of two is the same rescaling as for the two-copy CFI. This is necessary, otherwise the same measurement repeated side by side on independent copies of the system would give a measurement error for the complete system which is at least a factor of two greater than the measurement error for the individual system. 


It has been known for some time that the original formulation of the uncertainty principle was not a tight bound, and indeed Heisenberg's uncertainty principle has been violated experimentally~\cite{erhart2012experimental,rozema2012violation,demirel2016experimental}. However, prior to this work, it was not demonstrated, theoretically or experimentally, that the universally valid uncertainty relations which succeeded Heisenberg's uncertainty principle, could be violated. The metrological bound derived by Lu and Wang~\cite{lu2021incorporating} built on several previous uncertainty relations~\cite{ozawa2003universally,ozawa2004uncertainty,branciard2013error,ozawa2014error}. There is nothing incorrect in any of these works, however they all rely on a particular assumption from Ozawa's paper. In Ref.~\cite{ozawa2004uncertainty} Ozawa states \textquote{We assume that any joint measurements are carried out on single systems}. So it is built into the definition of all of these uncertainty relations that only separable measurements are considered. As collective measurements offer no advantage over separable measurements for pure states, we can expect the above uncertainty relations to hold for pure states. 


 \begin{figure}[t]
\includegraphics[width=0.5\textwidth]{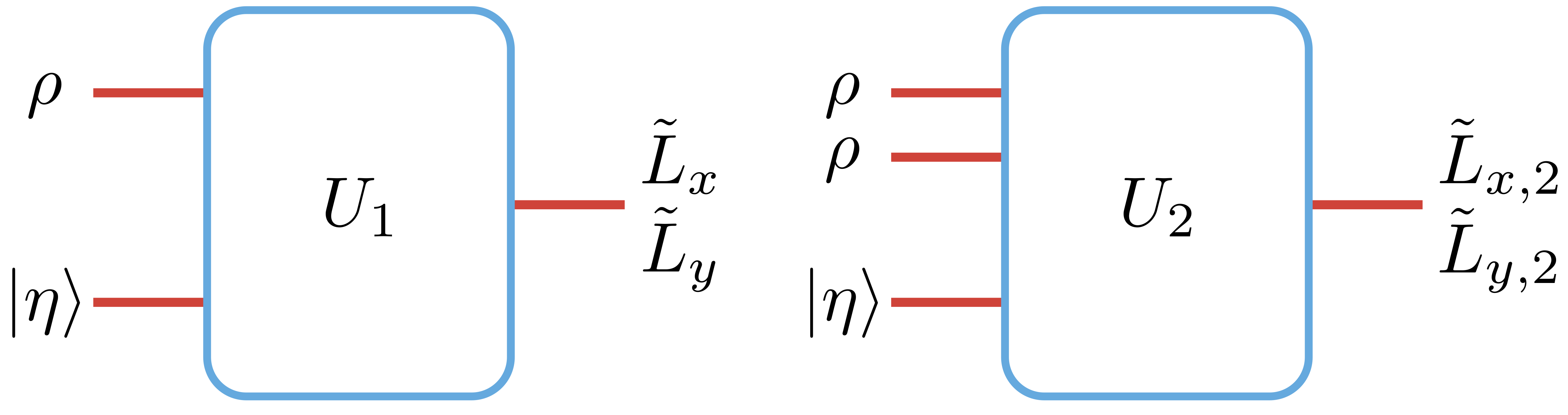}
\caption{\textbf{Schematic for estimating SLD operators.} As the SLD operators for this problem do not commute, it is necessary to measure an approximate version of these operators on an extended Hilbert space. $\rho$ is the input state and $\ket{\eta}$ is any ancilla state. The unitary matrices to be implemented are those that satisfy Eq.~\eqref{eq:unitcond}. The two-copy measurement is able to estimate the two-copy SLD operators better than what is allowed by the uncertainty relation, Eq.~\eqref{eq:ucrelation}, after accounting for a factor of two rescaling. $\tilde{L}_{x(y)}$ and $\tilde{L}_{x(y),2}$ are the approximate observables measured in the single- and two-copy schemes respectively.}
\label{fig:LWscheme}
\end{figure}

\subsection{Adjusting Lu and Wang's bound to allow for collective measurements}
\label{sn:adjustLW}
Eq.~(19) of the supplemental material of Lu and Wang's paper, Ref.~\cite{lu2021incorporating}, reads
\begin{equation}
\label{eq:LWregret}
R_{jj}\mathcal{F}_{kk}+R_{kk}\mathcal{F}_{jj}+2\sqrt{\mathcal{F}_{jj}\mathcal{F}_{kk}-D_{jk}^2}\sqrt{R_{jj}R_{kk}}\geq D_{jk}^2\;,
\end{equation}
where we have replaced $C_{jk}^2$ with $D_{jk}^2$, which strengthens the inequality. This bound holds when the regret is the difference between the SLD Fisher information and a separable measurement precision, $R_{jj}=\mathcal{F}_{jj}-F_{jj}^{\text{sep}}$. However, in reality when considering collective measurement precisions the regret can be reduced. By examining how much the regret can decrease when allowing for collective measurements, the LW uncertainty relation can be altered to accounts for collective measurements. We denote the CFI for the optimal collective measurement, i.e. a collective measurement on infinitely many copies of the probe state, as $F_{jj}^{\text{col}}$. Then the regret $R_{jj}$ can be reduced a factor $S_{jj}$ where
\begin{equation}
S_{jj}=\frac{\mathcal{F}_{jj}-F_{jj}^{\text{col}}}{\mathcal{F}_{jj}-F_{jj}^{\text{sep}}}\leq 1\;.
\end{equation}
We can therefore modify Eq.~\eqref{eq:LWregret} in the following way to account for collective measurements
\begin{equation}
\label{eq:LWregretmod}
R_{jj}\mathcal{F}_{kk}+R_{kk}\mathcal{F}_{jj}+2\sqrt{\mathcal{F}_{jj}\mathcal{F}_{kk}-D_{jk}^2}\sqrt{R_{jj}R_{kk}}\geq D_{jk}^2\times\min(S_{jj},S_{kk})\;.
\end{equation}
Unfortunately, $S_{jj}$ is not easily computed as there is no known way to find $F_{jj}^{\text{col}}$. Nevertheless, there are still situations where Eq.~\eqref{eq:LWregretmod} will be useful. For symmetric problems we have that $F_{jj}^{\text{col}}=F_{kk}^{\text{col}}=2/\mathcal{H}$. As the Holevo bound can be computed efficiently, $\min(S_{jj},S_{kk})$ can be computed efficiently in this case. 

%


\subsection{Probability simplex}
\label{SN:probsimp}
\begin{figure}[t]
\includegraphics[width=0.5\textwidth]{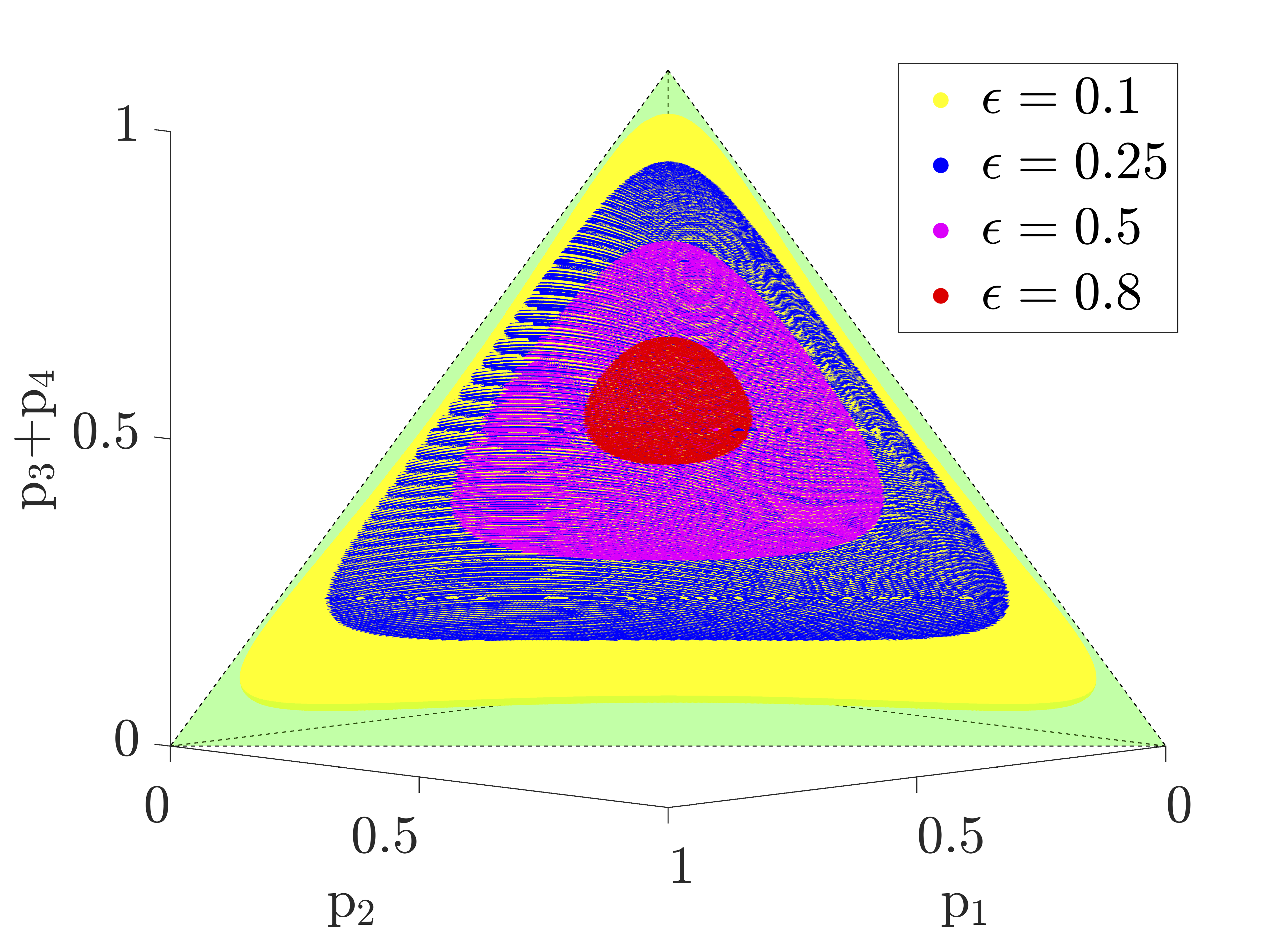}
\caption{\textbf{Probability simplex for the problem being considered.} Varying $\theta_x$ and $\theta_y$ generates a space known as a probability simplex. The green tetrahedron is the set of allowed probabilities. As there are only four measurement outcomes, the probability simplexes generated are restricted to lie on the surface of the tetrahedron. The probability simplexes decrease in size as $\epsilon$ increases which accounts for the increasing variance.}
\label{fig:simplex}
\end{figure}

In order to extract information about $\theta$, varying $\theta$ must change the probability of the measurement outcomes obtained. The different possible probability distributions form a space known as a probability simplex~\cite{sidhu2020geometric}. In Fig~\ref{fig:simplex} we plot the the probability simplex generated by varying $\theta_x$ and $\theta_y$ in the region $0$ to $2\pi$, as a geometrical interpretation of our two-copy measurement. As we can only plot a 3D simplex we combine two of the outcome probabilities into one axis. The three different axes of our simplex are $p_1$, $p_2$ and $p_3+p_4$, where $p_i$ is the probability of obtaining the $i$th measurement outcome from Eq.~\eqref{twocopyproj}. The probability simplex is shown for different values of $\epsilon$. When $\epsilon$ increases the area occupied by the simplex decreases, meaning neighbouring states are harder to distinguish. As $\epsilon\rightarrow1$ the simplex shrinks to the point where all four measurement outcomes are equally likely for all values of $\theta$. At this point it is impossible to discern any information about $\theta$ and so the variance goes to infinity.

\bibliography{col_meas_bibS}
\bibliographystyle{naturemag}